\documentclass[12pt,epsf,amssymb]{article}

\usepackage{graphicx}
\usepackage{amsfonts}
\usepackage{amsmath}
\usepackage[usenames]{color}
\usepackage{amssymb}
\usepackage[mathscr]{eucal} 
\usepackage{url}

\def\N{\mathbb{N}}
\def\Z{\mathbb{Z}}
\def\Q{\mathbb{Q}}
\def\R{\mathbb{R}}
\def\C{\mathbb{C}}
\def\P{\mathbb{P}}

\usepackage{here}

\usepackage{amsthm}
\theoremstyle{definition}

\begin{document}

\baselineskip 0.6cm
\newcommand{\vev}[1]{ \left\langle {#1} \right\rangle }
\newcommand{\bra}[1]{ \langle {#1} | }
\newcommand{\ket}[1]{ | {#1} \rangle }
\newcommand{\Dsl}{\mbox{\ooalign{\hfil/\hfil\crcr$D$}}}
\newcommand{\nequiv}{\mbox{\ooalign{\hfil/\hfil\crcr$\equiv$}}}
\newcommand{\nsupset}{\mbox{\ooalign{\hfil/\hfil\crcr$\supset$}}}
\newcommand{\nni}{\mbox{\ooalign{\hfil/\hfil\crcr$\ni$}}}
\newcommand{\nin}{\mbox{\ooalign{\hfil/\hfil\crcr$\in$}}}
\newcommand{\Slash}[1]{{\ooalign{\hfil/\hfil\crcr$#1$}}}
\newcommand{\EV}{ {\rm eV} }
\newcommand{\KEV}{ {\rm keV} }
\newcommand{\MEV}{ {\rm MeV} }
\newcommand{\GEV}{ {\rm GeV} }
\newcommand{\TEV}{ {\rm TeV} }

\def\diag{\mathop{\rm diag}\nolimits}
\def\tr{\mathop{\rm tr}}

\def\Spin{\mathop{\rm Spin}}
\def\SO{\mathop{\rm SO}}
\def\SU{\mathop{\rm SU}}
\def\U{\mathop{\rm U}}
\def\Sp{\mathop{\rm Sp}}
\def\SL{\mathop{\rm SL}}

\def\change#1#2{{\color{blue}#1}{\color{red} [#2]}\color{black}\hbox{}}

 \begin{titlepage}
  
 \begin{flushright}
 IPMU21-0071
 \end{flushright}

 \vskip 1cm
 \begin{center}
  
{\large \bf Direct Computation of Monodromy Matrices and \\
Classification of 4d ${\cal N}=2$ Heterotic--IIA Dual Vacua}
 
 \vskip 1.2cm

Yuichi Enoki and  Taizan Watari
  
 \vskip 0.4cm
  
  {\it 
     Kavli Institute for the Physics and Mathematics of the Universe (WPI), 
    the University of Tokyo, Kashiwa-no-ha 5-1-5, 277-8583, Japan
   }
 
 \vskip 1.5cm
    
 \abstract{We compute the monodromy matrices on the special geometry of 4d ${\cal N}=2$ Heterotic--IIA dual vacua in some simple cases by numerical evaluation of the period integrals, without assuming geometric background. The integrality of the monodromy matrices constrains some classification invariants of the string vacua. We also mention some mathematical open problems on period polynomials for modular form with poles.} 
 \end{center}
 \end{titlepage}
 
\tableofcontents

\newpage

\section{Introduction}
\label{sec:Intro}

It is a hard task to classify string vacua, or to classify 
modular invariant superconformal field theories (SCFT's) on worldsheet. 
Classification of Calabi--Yau manifolds will do a partial job, 
but it is also a hard task to classify geometry; explicit 
construction of geometry one by one in a certain method (e.g., complete 
intersections in toric varieties) does not tell us how many 
other Calabi--Yau manifolds are overlooked by that method. Furthermore, 
not all the SCFT's may be associated with non-linear sigma models 
of Calabi--Yau manifolds, in principle.  

In this article, we focus on string vacua with $\SO(3,1)$ 
Lorentz symmetry and ${\cal N}=2$ supersymmetry that have 
descriptions both by 
Heterotic string and Type IIA string theory \cite{KV}, and report a progress 
on the question above. The moduli space 
of such 4d ${\cal N}=2$ vacua forms a network of branches 
connected by the Coulomb--Higgs transitions, and individual branches 
are assigned invariants for classification: a pair of lattices 
$(\Lambda_S, \Lambda_T)$ and a finite number of integers, to be 
explained in the main text.  We will use facts that are known 
since 90's, and obtain constraints on those integer parameters. 

Here is a little more words on the idea. It has been known 
since 90's for a mirror pair of Type IIA compactification 
on a Calabi--Yau threefold $M$ and Type IIB compactification 
on a Calabi--Yau threefold $W$, that there is a monodromy 
of a symplectic projective section on the vector-multiplet 
moduli space; the monodromy matrices take value in the 
group of integer-valued symplectic transformations on $H^3(W;\Z)$. 
Monodromy matrices have been computed explicitly for 
explicitly constructed and chosen mirror manifolds $W$. 
There should exist, however, the notion of monodromy and its appropriate 
matrix representation, regardless of whether a given branch of 
4d ${\cal N}=2$ compactification is given by a non-linear sigma 
model of a Calabi--Yau threefold. In fact, with a careful reading 
of references such as \cite{deWit, AFGNT, KapLusThs, AP} and also 
\cite{AGNT-threshold, AGNT-95-high-deriv}, 
one will find that, for certain class of lattices $\Lambda_S$, 
the monodromy matrices can be computed from the abstract data of 
$(\Lambda_S,\Lambda_T)$ and the integers, without knowing whether 
a mirror non-linear sigma model exists.  
We require that the monodromy matrices should be integer-valued, 
and derive constraints on those integers for classification.\footnote{
The study in this article was inspired by positive evidence 
in \cite{ESW20, STW} that this requirement yields non-trivial 
conditions on the integers for classification. 
} %

In this article, we work on two cases, $\Lambda_S = \vev{+2}$ and 
$\Lambda_S = U$, as a proof of concept. In both cases, it turns out 
that each one of the branches of the moduli space has a region described 
by a non-linear sigma model; on the way to establish 
this claim, we have confirmed that two independent ways to read out 
the second Chern class of the target manifold $M$ yield consistent 
result. In the most non-trivial part of computing the monodromy matrices, 
we evaluated numerically the period polynomials of the automorphic forms 
of the isometry groups associated with the lattices $\Lambda_S$. 

Sections \ref{ssec:prelim}--\ref{ssec:rho=1-review} should be regarded 
as a process of reading the literatures; technical materials in this part 
are entirely from references, except a few minor clarifications 
(footnotes \ref{fn:factor2-vs-AFGNT} and \ref{fn:KG-Phi-relatn}). 
Technical analysis along the idea written in the middle of 
section \ref{ssec:monodromy} will start at the end of 
section \ref{sssec:deg2-review}. Some useful facts on modular forms 
are summarized in the appendix \ref{sec:mod-form}. In the appendix 
 \ref{sec:Eichler-coh}, we will make a brief comment on how the 
coarse and fine classifications in \cite{EW19} are related to 
the Eichler cohomology.

\section{Monodromy of Projective Symplectic Sections}

\subsection{Preliminaries}
\label{ssec:prelim}

{\bf The framework in the Heterotic string language:}

In a Heterotic string compactification with the $\SO(3,1)$ 
Lorentz symmetry, ${\cal N}=2$ spacetime supersymmetry implies 
special features \cite{BD} in the worldsheet SCFT with 
central charge $(c,\tilde{c}) = (22,9)$. The right-moving sector 
contains a $\tilde{c}=3$ algebra of two free bosons and two free fermions
forming a system with $N=2$ superconformal symmetry, and a $\tilde{c}=6$ 
$N=4$ superconformal algebra. In the left-moving sector, let $\rho$ be 
the number of free chiral bosons; there is also a Virasoro algebra 
of $c=22-\rho$. The total Hilbert space in the Ramond sector 
has the structure of 
\begin{align}
 {\cal H}^{\rm tot}_R = \oplus_{\gamma \in G_S} {\cal H}^{(\rho,3)}_\gamma \otimes 
   \left( \oplus_{(\tilde{h},\tilde{I})} 
      {\cal H}^{(22-\rho,0)}_{\gamma,(\tilde{h},\tilde{I})} \otimes 
      \widetilde{\cal H}^{(0,6)}_{(\tilde{h},\tilde{I})} \right), 
\end{align}
with each factor forming a representation space of the algebras of 
$(c,\tilde{c})=(\rho,3)$, $(22-\rho,0)$ and $(0,6)$, respectively; 
irreducible Ramond-type representations of the $\tilde{c}=6$ $N=4$ 
superconformal algebra are labeled by $(\tilde{h},\tilde{I})$. 
The group $G_S$ and its elements $\gamma$ are explained shortly. 
The spectrum is assumed to yield a modular invariant partition 
function; spacetime ($\R^{3,1}$) filling NS5-branes are assumed 
to be absent. 

The $(c,\tilde{c})=(\rho,3)$ sector is the $N=(0,1)$ supersymmetrization of 
a lattice CFT, where the lattice---denoted by $\widetilde{\Lambda}_S$---is 
even and of signature $(2,\rho)$. In this article, we will consider the 
cases where the lattice $\widetilde{\Lambda}_S$ has a structure of 
$\widetilde{\Lambda}_S = U[-1] \oplus \Lambda_S$ for an even lattice\footnote{
An even unimodular (self-dual) lattice of signature $(r_+, r_-)$ is 
denoted by ${\rm II}_{r_+,r_-}$. The lattice ${\rm II}_{1,1}$ is also 
denoted by $U$. For a lattice $L$, $L[n]$ denotes the lattice 
that is isomorphic to $L$ as a free abelian group, and whose bilinear 
form (intersection form) is $n$ times that of $L$.  
} %
$\Lambda_S$ of signature $(1, \rho-1)$, and also has a primitive embedding 
$\widetilde{\Lambda}_S \hookrightarrow {\rm II}_{4,20}$. 
The orthogonal complement of $\widetilde{\Lambda}_S$ within ${\rm II}_{4,20}$
is denoted by $\Lambda_T$. The discriminant group and the quadratic 
discriminant form of the lattice $\Lambda_S$ are denoted by $G_S$ and $q_S$, 
respectively.  

Besides the lattice pair $(\widetilde{\Lambda}_S, \Lambda_T)$, the 
generating functions of indices (elliptic genus)
\begin{align}
 \frac{\Phi_\gamma}{\eta^{22-\rho}}(q)  := {\rm Tr}_{ \oplus_{(\tilde{h},\tilde{I})} 
     \left( {\cal H}^{(22-\rho,0)}_{\gamma,(\tilde{h},\tilde{I})} \otimes 
     \widetilde{\cal H}^{(0,6)}_{(\tilde{h},\tilde{I})}  \right)} \left[ 
    q^{L_0-\frac{22-\rho}{24}} \bar{q}^{\tilde{L}_0-\frac{6}{24}} (-1)^{F_R} \right]
\end{align}
serve as classification invariants of individual branches of 4d ${\cal N}=2$ 
moduli space of Heterotic string compactifications. The 
set of generating functions $\{ \Phi_\gamma \; | \; \gamma \in G_S\}$ 
is a vector-valued modular form of weight $(11-\rho/2)$ associated with 
the quadratic discriminant form $(G_S, q_S)$ derived from the lattice 
$\Lambda_S$. The modular nature implies that the whole $\{\Phi_\gamma \}$
is completely fixed already, when all the first Fourier coefficients of 
$\Phi_\gamma$'s---$\Phi_\gamma = n_\gamma q^{\nu_\gamma} + \cdots $ where 
$0 \leq \nu_\gamma < 1$---are specified. For this reason, for a given 
$\Lambda_S$, the vector-valued modular form $\{\Phi_\gamma \}$ contains 
only a finite number of $\Z$-valued classification invariants of the 
branches of the moduli space. 

It has been observed that a little more classification invariants 
are necessary in distinguishing branches of moduli space 
in addition to $(\widetilde{\Lambda}_S,\Lambda_T)$ and $\{ \Phi_\gamma \}$. 
Reference \cite{EW19} introduced yet another vector-valued modular form 
$\{ \Psi_\gamma \; | \; \gamma \in G_S\}$ of weight-$(13-\rho/2)$ 
to serve the purpose by following the idea in \cite{HM, Stieberger}. 
The modular form $\{ \Psi_\gamma \}$ is also parametrized by a finite 
number of integers; more information on these integer parameters 
will be provided later in this article, when it becomes necessary
((\ref{eq:para-in-Psi-deg2}) and (\ref{eq:para-in-Psi-U})).  

{\bf Prepotentials:}

The 4d field theory description of the Heterotic string compactification 
in question has $(\rho+2)$ abelian vector bosons. Electric charges 
are in $\widetilde{\Lambda}_S^\vee$, so the magnetic charges 
are in $\widetilde{\Lambda}_S$. 
The mass of BPS states of the 4d ${\cal N}=2$ spacetime supersymmetry 
is given by
\begin{align}
  m & \; = |Z|/\sqrt{G_N}, \qquad \quad Z = (v_I X^I + m^I F_I) e^{\hat{K}/2}, 
     \label{eq:4dSUSY-central-C} \\
  & \hat{K} = - \ln \left( i (X^I \overline{F}_I - \overline{X}^I F_I) \right)
\end{align}
for some projective symplectic section $\Pi = (X^I, F_I)^T$, where 
$\nu = (v, m) = (v_I, m^I)$ is the electric and magnetic charges 
of a BPS state under the $(\rho+2)$ U(1) vector bosons.  
The 4d Newton constant is denoted by $G_N$. Electric charges 
$v \in \widetilde{\Lambda}_S^\vee$ [resp. magnetic charges 
$m \in \widetilde{\Lambda}_S$] are described as $v = e^I v_I$, $v_I \in \Z$
[resp. $m = e_I m^I$, $m^I \in \Z$] by choosing a basis\footnote{
The intersection form of $\widetilde{\Lambda}_S$ [resp. $\Lambda_S$] 
is denoted by $\widetilde{C}_{IJ}$ [resp. $C_{ab}$], where 
$I,J = 0,\sharp, 1, \cdots, \rho$ [resp. $a, b = 1,\cdots, \rho$]. 
The lattice $U[-1]$ has the intersection form $(e_0, e_0) = 
(e_\sharp, e_\sharp) = 0$ and $(e_0, e_\sharp) = -1$. 
} %
$\{ e^{I = 0,\sharp, 1,\cdots, \rho} \}$ of $\widetilde{\Lambda}_S^\vee$
[resp. $\{ e_{I=0,\sharp,1,\cdots, \rho} \}$ of $\widetilde{\Lambda}_S$]. 
So, relative normalization between the electric part $(X^I)$ and the 
magnetic part $(F_I)$ of the section $\Pi = (X^I, F_I)^T$ 
is no longer arbitrary. 

For a given branch of moduli space, the projective symplectic section 
$\Pi = (X^I, F_I)^T$ is locally a function of flat coordinates 
$(s, t^{a=1,\cdots, \rho})$ of the vector-multiplet moduli space; we will 
provide more information on the coordinates $t^{a=1,\cdots, \rho}$ in 
section \ref{ssec:monodromy}, but we note for the moment that 
$t = e_a t^a$ can be regarded as an element of $\Lambda_S \otimes \C$; 
the coordinate $s$ is normalized 
so that the gauge coupling constant of a 4d non-abelian gauge field 
from a level-$k$ current algebra in Heterotic string is given by 
$(4\pi/g_{YM}^2) = k {\rm Im}(s)$. The section $\Pi$ is of the form\footnote{
Already the relative normalization between $X^I$'s and $F_I$'s 
has been fixed here, which means that the normalization of ${\cal F}$
is also fixed. Now, there is no ambiguity in writing down the 
gauge kinetic term in a 4d ${\cal N}=2$ supergravity in the convention 
adopted in this article. \\ 
Let the covariant derivatives on purely electrically charged states 
be $\nabla = d -i A^I v_I$ with $I \in \{ 0,\sharp, 1,\cdots, \rho\}$, 
and $F^I_{\mu\nu} := \partial_\mu A^I_\nu - \partial_\nu A^I_\mu$. Then 
\begin{align}
  {\cal L} \supset - \frac{1}{4}\frac{{\rm Im}({\cal N}_{IJ})}{2\pi} 
      F_{\mu \nu}^I F^{J\mu\nu};  
   \label{eq:gauge-kin-1}
\end{align}
the matrix ${\cal N}_{IJ}$ is obtained by the electro-magnetic dual 
transformation (linear fractional transformation) 
as in \cite{Ceresole-dualitytransformatn} from ${\cal N}^{(A)}_{IJ}$, where 
\begin{align}
 \overline{\cal N}^{(A)}_{IJ} = F_{IJ} -2i \frac{{\rm Im}(F_{IK})\overline{X}^K \; {\rm Im}(F_{JL}) \overline{X}^L }{{\rm Im}(F_{PQ}) \overline{X}^P \overline{X}^Q}, \label{eq:gauge-kin-2}
\end{align}
using $F(X) := (X^0)^2 {\cal F}(t^i =: X^i/X^0)$, $F_I = \partial_{X^I}F$
and $F_{IJ} = \partial_{X^I}\partial_{X^J}F$. 
} %
\begin{align}
  \Pi = \Pi^{\cal F} & \; = (X^{I}, F_I)^T \\
  & \; =: ( 1, (\partial_s {\cal F}), t^a, \; 
     (2 - t\partial_t -s\partial_s){\cal F}, -s, (\partial_{t^a}{\cal F}) )^T 
    \label{eq:periods-Het-weak} 
\end{align}
for a prepotential ${\cal F}$ of the given branch; the 
prepotential ${\cal F}(s,t)$ is of the form 
\begin{align}
 {\cal F} & \; = \frac{s}{2}(t,t) + \frac{d_{abc}}{3!} t^a t^b t^c 
   - \frac{\zeta(3)}{(2\pi i)^3}\frac{\chi}{2}
   + \frac{1}{(2\pi i)^3} \sum_{k \in \N_{\geq 0}} 
   \sum_{w \in \Lambda_{S}^\vee}^{({\rm Im}(\vev{w,t})\geq 0)} n_{w,k} 
       {\rm Li}_3(e^{2\pi i k s} e^{2\pi i \vev{w, t}})  
        \nonumber \\
   & \;   - \frac{a_{ab}}{2}t^at^b - \frac{b_a}{24} t^a, 
   \label{eq:prepot-Het-weak}
\end{align}
for some appropriately chosen parameters $d_{abc}$, $a_{ab}$, $b_a$, 
$\chi$ and $n_{w,k}$ where $a,b,c =1,\cdots,\rho$. 
Rationale for the $s$ and $t$-dependence in (\ref{eq:periods-Het-weak}) 
is written, for example, in the appendix of \cite{ESW20}
(reasonings in both Heterotic and Type IIA perspectives are used).
Note that there is no ambiguity left for the normalization of ${\cal F}$, 
and hence of those parameters. 

A prepotential itself is not a physical observable, although 
the spectrum of 4d ${\cal N}=2$ BPS states is. The spectrum 
remains unchanged under a symplectic transformation on 
$\Pi^{\cal F}= (X^I, F_I)^T$ and its conjugate action on the charge 
$\nu = (v_I, m^I)$; when the symplectic transformation 
is in a certain subgroup\footnote{
\label{fn:symp-frame-change}
In the notation to be introduced in (\ref{eq:notatn-M(g,L)}), 
$\Pi' = M({\bf 1}, \Lambda) \Pi$ 
with $\Lambda_{00} = \Lambda_{\sharp \sharp}
= \Lambda_{\sharp 0} = \Lambda_{0 \sharp} = 0$. 
Such $2(\rho+2) \times 2(\rho+2)$ matrices form a subgroup of 
${\rm Sp}(2(\rho+2);\Z)$. 
The integers $\delta n_a$ and $\Lambda'_{0a}$ in the main text 
correspond to $\Lambda_{\sharp a}$ and $\Lambda_{0a}+\Lambda_{\sharp a}$, 
respectively. \\
Depending on whether we use the section $\Pi^{\cal F}$ or $\Pi^{{\cal F}'}$, 
monodromy matrices to be computed in this article become either 
$M_{\tilde{g}}$ or $M({\bf 1}, \Lambda) \cdot M_{\tilde{g}} \cdot 
M({\bf 1}, -\Lambda)$. 
 } %
 of ${\rm Sp}(2(\rho+2);\Z)$,  
the section $\Pi'$ after the transformation may still be fitted 
by (\ref{eq:periods-Het-weak}) for some prepotential ${\cal F}'$ of 
the form (\ref{eq:prepot-Het-weak}), $\Pi' = \Pi^{{\cal F}'}$,  
but with parameters $d'_{abc}$, $a'_{ab}$, $b'_a$ different 
from the parameters before the transformation;  
\begin{align}
  d'_{abc} -d_{abc} = [(\delta n)_a C_{bc} + {\rm cyclic} ], \quad 
  a'_{ab} -a_{ab} = - \Lambda_{ab} , \quad
  b'_a  - b_a = 24 (\delta n_a) -24 \Lambda'_{0a} 
  \label{eq:abd-ambiguity}
\end{align}
$(\delta n)_a, \Lambda_{ab}, \Lambda'_{0a} \in \Z$. 

In the branch of 4d ${\cal N}=2$ moduli space with classification 
invariants $(\Lambda_S, \Lambda_T)$, $\{ \Phi_\gamma \}$ and $\{ \Psi_\gamma \}$, 
some of the parameters in the prepotential are determined by \cite{HM}
\begin{align}
   n_{w, k=0} = c_{[w]}((w,w)/2) \quad 
       (w \in \Lambda_S^\vee, \; {\rm Im}(\vev{w,t}) > 0), 
    \qquad  \quad \chi = - c_0(0),   
  \label{eq:matching-GV+chi}
\end{align}
where the Fourier expansion coefficients of $\{ \Phi_\gamma \}$, 
\begin{align}
 F_\gamma := \frac{\Phi_\gamma}{\eta^{24}}(\tau) =: \sum_{\nu} c_\gamma(\nu) q^\nu 
\end{align}
are used. 
This is done by computing Heterotic string genus-1 (1-loop) corrections 
to the holomorphic $R^2$ (gravitational) term and the probe gauge group 
kinetic function, and matching the results to the parameters in the 
effective theory on $\R^{3,1}$ (see e.g., \cite{DKL, HM, CurioLust-etal, 
HM-grav, Stieberger}). 
The same procedure also determines the parameters $d_{abc}$ 
 in terms of $(\Lambda_S, \Lambda_T)$, $\{ \Phi_\gamma \}$ 
and $\{ \Psi_\gamma \}$ modulo the ambiguity in (\ref{eq:abd-ambiguity}),  
but is able to constrain $a_{ab}$ and $b_a$'s only to the extent that 
\begin{align}
   a_{ab}, \quad b_a \in \R.  
  \label{eq:matching-on-aNb}
\end{align}
So, at this moment, we should think that $(a_{ab})_{+\Z} \in \R/\Z$ and 
$(b_a)_{+24\Z} \in \R/24\Z$ are also invariants characterizing the branch 
in addition to $(\Lambda_S, \Lambda_T)$, $\{ \Phi_\gamma \}$ and 
$\{ \Psi_\gamma \}$. 

For a choice of $(\Lambda_S, \Lambda_T)$, it is possible to work out 
a finite number of independent integer parameters that specify 
the vector-valued modular forms $\{ \Phi_\gamma \}$ and $\{ \Psi_\gamma \}$ 
completely. Those independent integer parameters are further subject to 
some number of inequalities, as discussed in \cite{EW19}.  
Starting from section \ref{ssec:monodromy}, we derive additional 
consistency conditions on those independent integer parameters, 
$(a_{ab})_{+\Z}$ and $(b_a)_{+24\Z}$.  

{\bf Type IIA non-linear sigma model phase:}

Such data as $\Lambda_S \hookrightarrow {\rm II}_{3,19}$, 
$\{\Phi_\gamma \}$ and $\{ \Psi_\gamma \}$ can also be characterized 
in the language of Type IIA string compactification.
Certainly the dictionary of the Heterotic--Type IIA duality has been 
studied and verified mostly in cases the Type IIA description 
is given by a Calabi--Yau compactification, with a non-linear 
sigma model on worldsheet (cf \cite{HS-MSW-GSY-DM}). It is not difficult 
to fix the dictionary with a guess work,\footnote{
First, there must be a basis of Ramond--Ramond ground states that 
corresponds to a basis of the spacetime U(1) vector bosons in which 
charges $(v_I, m^I)$ are integers. Second, within the $ca$-ring states 
with the conformal weight $(1/2, 1/2)$, one may also choose a basis 
that are tied under a spectral flow with some $(\rho+1)$ elements 
in the basis of the Ramond--Ramond ground states. Let $\phi_{\sharp^\wedge}$
and $\phi_{a=1,\cdots, \rho}$ be those $ca$-ring states. Now, the data 
$C_{ab}$ (i.e., $\Lambda_S$) and the parameters $d_{abc}$ and $n_{w,k}$ 
are read out from the structure constants of the $ca$-ring. The class 
of Type IIA compactifications to be considered in this article is 
those with a state $\phi_{\sharp^\wedge}$ and the corresponding flat 
coordinate $s$ so that the three point functions are of the form 
$\vev{\phi_{\sharp^\wedge} \phi_{\sharp^\wedge} \phi_*}
= {\cal O}(e^{2\pi is})$ and 
$\vev{\phi_{\sharp^\wedge} \phi_a \phi_b} = C_{ab} + {\cal O}(e^{2\pi i s})$. \\
The vector-valued modular form $\{ \Phi_\gamma \}$ encodes the helicity 
supertrace of purely electrically charged BPS particles on $\R^{3,1}$
\cite{BSV-KKV}. 
} %
 even in cases the Type IIA description is not necessarily
associated with a Calabi--Yau-target non-linear sigma model.
For this reason, Type IIA compactifications that are not obviously 
related to a Calabi--Yau compactification are also within the framework 
of study in this article. 
 
If a branch of moduli space contains a phase given by $M$-target 
$N=(2,2)$ supersymmetric non-linear sigma model on worldsheet, where 
$M$ is a Calabi--Yau threefold, one may choose a basis 
$\{ D_{\sharp^\wedge}, D_{a=1,\cdots, \rho} \}$ of $H^2(M;\Z)$, with 
$D_{\sharp^\wedge}$ represented by the K3-fiber class \cite{KLM}. 
The cubic part of the prepotential ${\cal F}$ is the trilinear 
intersection form on $H^2(M;\C)$, 
\begin{align}
  (C_{ab}, d_{abc}):  H^2(M;\C) \ni J := s D_{\sharp^\wedge} + t^a D_a \longmapsto 
   \frac{1}{3!} \int_M J \wedge J \wedge J 
   = \frac{s}{2}C_{ab}t^at^b+\frac{d_{abc}}{3!}t^a t^b t^c, 
  \label{eq:def-geom-trilin}
\end{align}
and the holomorphic $R^2$ term in the 4d action 
(\cite{AGNT-topological}, \cite[\S8]{BCOV-2}) 
involves the information \cite{BCOV-1} 
\begin{align}
 (24, (c_2)_a): H^2(M;\C) \ni J \longmapsto
    \int_M c_2(TM) \wedge J = 24 s + (c_2)_a t^a. 
   \label{eq:def-geom-c2}
\end{align}
For a different choice of a basis $\{ D_{\sharp^\wedge}, 
D'_a := D_a + (\delta n_a) D_{\sharp^\wedge} \}$, 
the same element $J \in H^2(M;\C)$ is regarded as 
$s' D_{\sharp^\wedge} + t^a D'_a$ with $s' = s-(\delta n_a) t^a$; 
the topological numbers $\int_M D'_a \cdot D'_b \cdot D'_c =: d'_{abc}$ 
[resp. $\int_M c_2(TM) \cdot D'_a =: (c_2')_a$] are different from 
$d_{abc}$ [resp. $(c_2)_a$] by $+\{(\delta n)_a C_{bc} + {\rm cyclic}\}$
[resp. $+24 (\delta n)_a$]. The symplectic transformations 
in (\ref{eq:abd-ambiguity}) parametrized by $\delta n_a$ correspond to 
this change of basis. 
The symplectic transformations in (\ref{eq:abd-ambiguity}) parametrized 
by $\Lambda_{ab}$ and $\Lambda'_{0a}$ just correspond to choosing different 
basis elements of magnetically charged states. 

It is known in a Type IIA vacuum given by an $M$-target non-linear 
sigma model that \cite{open-Z-basis} 
\begin{align}
 (a_{ab})_{+\Z} = \frac{1}{2} D_a \cdot D_a \cdot D_b + \Z
  = \frac{1}{2} D_a \cdot D_b \cdot D_a + \Z 
  \label{eq:open-geom-cond-a}
\end{align}
and 
\begin{align}
  (24\Z) s + (b_a)_{+24\Z} t^a = 
   \int_M \left( c_2(TM) + 24 H^4(M;\Z) \right) \wedge J. 
   \label{eq:open-geom-cond-b}
\end{align}
The latter property implies $(b_a)_{+24\Z} \equiv (c_2)_a + 24\Z$. 
In this article, however, we do not assume 
that (\ref{eq:open-geom-cond-a}, \ref{eq:open-geom-cond-b}) are satisfied 
from the beginning, because there may be a branch of moduli space 
of Type IIA vacua that are not associated with any 
Calabi--Yau compactification. As a result of the monodromy analysis 
in this article, however, we will see for some lattices $\Lambda_S$
that both (\ref{eq:open-geom-cond-a}, \ref{eq:open-geom-cond-b}) 
are always satisfied. 

\subsection{Monodromy within the Heterotic-perturbative Region}
\label{ssec:monodromy}

The vector-multiplet moduli space of a given branch is known to have
the following approximate form 
\begin{align}
  \left( ({\cal H}_{s; {\rm Im}(s)\gg 1}/\Z D) \times 
         D(\widetilde{\Lambda}_S) \right) / [\Gamma_S] = 
  \left( {\cal H}_{s, {\rm Im}(s) \gg 1} \times D(\widetilde{\Lambda}_S) \right)
   / \Z D \times [\Gamma_S]
\label{eq:vect-moduli-sp-apprx}
\end{align}
in the region ${\rm Im}(s) \gg 1$ (the perturbative region 
in the Heterotic string language). Here, ${\cal H}_s$ is the complex 
upper half-plane parametrized by $s$, and 
\begin{align}
  D(\widetilde{\Lambda}_S ) := \P \left\{ 
    \mho \in \widetilde{\Lambda}_S \otimes \C \; | \; (\mho, \mho) = 0, \; 
     (\mho, \overline{\mho}) > 0 \right\},   
\end{align}
for which one can use the following parametrization:
\begin{align}
 \mho(t) = e_0 + e_{\sharp} \frac{(t,t)}{2} + e_a t^a, \qquad 
    t := e_a t^a \in \Lambda_S \otimes \C, 
\end{align}
using the basis $\{ e_0, e_\sharp, e_{a=1,\cdots, \rho} \}$ of 
$\widetilde{\Lambda}_S$. In the region ${\rm Im}(s) \gg 1$, we have 
chosen in eq. (\ref{eq:periods-Het-weak}) the electric part of the 
projective symplectic section $\Pi$ as 
\begin{align}
(X^I ) = (\mho^I(t)) + {\cal O}(e^{2\pi i s}), \qquad 
  (\mho^I(t)) = \left( 1, \frac{(t,t)}{2}, t^a \right)^T.
\end{align}

Within the quotient group, a generator $D$ acts only on ${\cal H}_s$, 
and it does as $s \rightarrow s + 1$. The group $[\Gamma_S]$
is a subgroup of the lattice isometry group 
${\rm Isom}(\widetilde{\Lambda}_S)$ containing  
\begin{align}
\Gamma_S := {\rm Ker}\left[ {\rm Isom}(\widetilde{\Lambda}_S) \rightarrow 
{\rm Isom}(G_S,q_S) \right] .
\end{align}
The subgroup $[\Gamma_S]$ may be as large as the set of lattice 
isometries of $\widetilde{\Lambda}_S \subset {\rm II}_{4,20}$ that 
can be lifted to an isometry of ${\rm II}_{4,20}$ (cf \cite{STW}). 
In the cases to be worked out explicitly in this article, where 
$\Lambda_S = \vev{+2}$ and $\Lambda_S = U$, this ambiguity does not 
matter because all those subgroups agree with the entire 
${\rm Isom}(\widetilde{\Lambda}_S)$. 

Within the moduli space (\ref{eq:vect-moduli-sp-apprx}), there 
are complex codimension-1 loci where 4d effective field theory 
has extra massless states charged under the $(\rho+2)$ U(1) 
vector fields. At 
\begin{align}
  X(v) := \left\{ [\mho] \in D(\widetilde{\Lambda}_S) \; | \; 
     (v, \mho) = 0 \right\}  
\end{align}
for $v \in \widetilde{\Lambda}_S^\vee$ satisfying $-2 \leq v^2 < 0$, 
electrically charged states with $v' \in \Q v \cap \widetilde{\Lambda}_S^\vee$
and $-2 \leq (v')^2 < 0$ become massless, with the multiplicity 
governed by the classification invariants $n_{\gamma = [v]}$. Those light 
states give rise to logarithmic singularity in the gauge 
coupling constants of the $(\rho + 2)$ U(1) vector fields around 
\cite{AFGNT}
\begin{align}
 X_{\rm singl.} := \cup_{v \in \widetilde{\Lambda}_S^\vee; -2 \leq v^2 < 0} X(v) \subset 
  D(\widetilde{\Lambda}_S) .
\end{align}
So, the projective symplectic section $\Pi=(X^I, F_I)^T$ has non-trivial 
monodromy around the complex codimension-1 locus $X_{\rm singl.}$ in 
$D(\widetilde{\Lambda}_S)$. 

Fix a base point $(s_0, t_0)$ in 
\begin{align}
 {\cal H}_{s; {\rm Im}(s)\gg 1} \times 
   (D(\widetilde{\Lambda}_S) \backslash X_{\rm singl.})
  \label{eq:vect-moduli-sp-apprx-woX-noQnt}
\end{align}
and its image $[(s_0,t_0)]$ in the moduli 
space (\ref{eq:vect-moduli-sp-apprx}). For a loop $\gamma_{\tilde{g}}$ in 
\begin{align}
  \left( ({\cal H}_{s,{\rm Im}(s)\gg 1}/\Z D) \times 
        (D(\widetilde{\Lambda}_S) \backslash X_{\rm singl.}) \right) /[\Gamma_S]
   \label{eq:vect-moduli-sp-apprx-woX}
\end{align}
with the base point $[(s_0, t_0)]$, one may think of analytically 
continuing the section $\Pi^{\cal F}(s,t)$ of a fixed prepotential ${\cal F}$ 
along the path $\tilde{\gamma}_{\tilde{g}}$ 
in (\ref{eq:vect-moduli-sp-apprx-woX-noQnt}), the lift of $\gamma_{\tilde{g}}$. 
Then there must be a matrix $M_{\tilde{g}} \in {\rm Sp}(2(\rho+2);\Z)^{\rm H.el}$
so that\footnote{
Alternatively, one may assign a symplectic matrix $M_{\tilde{g}}$ for a loop 
$\gamma_{\tilde{g}}$ in (\ref{eq:vect-moduli-sp-apprx-woX}) by using 
analytic continuation of $\Pi$ in the reverse direction of $\gamma_{\tilde{g}}$, 
or placing the matrix $M_{\tilde{g}}$ in the left hand side of 
(\ref{eq:duality-map-proj-reltn}). In this article, we follow the 
assignment that looks popular in the literatures. In order for this 
assignment to be a homomorphism, we set the composition law 
of the loops as in (\ref{eq:path-composition-law-2}). 
See also the appendix \ref{sec:Eichler-coh}. 
} %
\begin{align}
(\Pi_{@ (s,t) \sim (s_0,t_0)}^{\cal F})^{{\rm contin.along~} \gamma_{\tilde{g}}} =_{\rm proj.} 
M_{\tilde{g}}  \cdot \Pi_{@ (s,t) \sim (s_0, t_0)}^{\cal F},
    \label{eq:duality-map-proj-reltn}
\end{align}
In a simplified notation, we may drop reference to a fixed ${\cal F}$, and 
write 
\begin{align}
   \Pi(s^{\tilde{g}}, t^{\tilde{g}}) =_{\rm proj.} M_{\tilde{g}} \cdot \Pi(s,t).  
\end{align}
The matrix $M_{\tilde{g}}$ has the form of $M(g, \Lambda_{\tilde{g}}) \in 
{\rm Sp}(2(\rho+2);\Z)^{\rm H.el}$, where \cite{Ceresole-dualitytransformatn}
\begin{align}
  {\rm Sp}(2(\rho+2);\Z)^{\rm H.el} & \; := 
    \left\{ \left. \left( \begin{array}{cc}  U & Z \\ W & V \end{array} \right)
      \in {\rm Sp}(2(\rho+2);\Z) \; \right| \; Z=0 \right\}, \\
  M(U, \Lambda) & \; := \left( \begin{array}{cc}
     U & 0 \\ V \cdot \Lambda & V \end{array} \right), 
   \qquad   U = (V^T)^{-1}, \quad \Lambda^T = \Lambda. 
   \label{eq:notatn-M(g,L)} 
\end{align}
The element $g$ in $M_{\tilde{g}} = M(g, \Lambda_{\tilde{g}})$
is the lattice isometry that maps the starting point 
$[\mho(t_0)]$ of the path $\tilde{\gamma}_{\tilde{g}}$ 
in the covering space $D(\widetilde{\Lambda}_S)$ to 
the endpoint $[\mho(t_0^{\tilde{g}})] \in D(\widetilde{\Lambda}_S)$.

For the assignment $\gamma_{\tilde{g}} \longmapsto M_{\tilde{g}}$ to be a 
homomorphism,\footnote{
\label{fn:-1-projectiveness}
Because the condition (\ref{eq:duality-map-proj-reltn}) leaves the 
freedom of changing the matrix $M_{\tilde{g}}$ by multiplying 
$- {\bf 1}_{2(\rho+2)\times 2(\rho+2)}$, we should expect that 
the assignment $\gamma_{\tilde{g}} \longmapsto M_{\tilde{g}} \in 
{\rm Sp}(2(\rho+2);\Z)^{\rm H.el}$ can be a homomorphism only projectively, 
with the fudge factor $- {\bf 1}_{2(\rho+2)\times 2(\rho+2)}$.  
That should be kept in mind when we 
talk of multiplying matrices $M_{\tilde{g}_a} \cdot M_{\tilde{g}_b}$, although 
we will use little space to mention this issue in the main text 
of this article. 
} %
\begin{align}
 \Pi(s^{(\tilde{g}_a \cdot \tilde{g}_b)}, t^{(\tilde{g}_a \cdot \tilde{g}_b)}) =_{\rm proj} 
   M_{\tilde{g}_a} \cdot M_{\tilde{g}_b} \cdot \Pi(s, t),  
 \label{eq:path-composition-law}
\end{align}
the path $\tilde{\gamma}_{(\tilde{g}_a \cdot \tilde{g}_b)}$ starting from 
$[\mho(t_0)] \in D(\widetilde{\Lambda}_S)$ has to end at 
$g_a \cdot (g_b \cdot [\mho(t_0)])$ in $D(\widetilde{\Lambda}_S)$. 
So, the composition law of the paths 
in (\ref{eq:vect-moduli-sp-apprx-woX-noQnt}) starting from $(s_0, t_0)$
[resp. the loops in (\ref{eq:vect-moduli-sp-apprx-woX}) with the base point 
$[(s_0, t_0)]$] should be 
\begin{align}
   \tilde{\gamma}_{(\tilde{g}_a \cdot \tilde{g}_b)} := 
    (\tilde{\gamma}_{\tilde{g}_b})^{g_a} \circ \tilde{\gamma}_{\tilde{g}_a}, \qquad 
 \left[ {\rm resp.} \quad 
   \gamma_{(\tilde{g}_a \cdot \tilde{g}_b)} := \gamma_{\tilde{g}_b} \circ 
\gamma_{\tilde{g}_a} \right]
   \label{eq:path-composition-law-2}
\end{align}
in the standard notation of the composition rule in fundamental groups. 

The monodromy matrices are completely understood for loops 
$\gamma_{\tilde{g}}$ in (\ref{eq:vect-moduli-sp-apprx-woX}) that are still 
regarded as loops when lifted to $({\cal H}_{s, {\rm Im}(s)\gg 1}/\Z D) \times 
(D(\widetilde{\Lambda}_S) \backslash X_{\rm singl.})$. All those monodromy 
matrices are of the form $M({\bf 1}, \Lambda_{\tilde{g}})$. For the loop 
$\gamma_D$ in the cylindrical fiber ${\cal H}_s /\Z D$, the monodromy matrix is 
\begin{align}
 M_D = M({\bf 1}, \widetilde{C}). 
\end{align}
For a loop $\tilde{\gamma}_{\tilde{g}(v)}$ that maintains constant $s=s_0$ and 
goes around $X(v) \subset D(\widetilde{\Lambda}_S)$ in the 
counter-clockwise direction (gaining phase $+2\pi$), 
the matrix representation is\footnote{
\label{fn:factor2-vs-AFGNT}
The dilaton superfield $S$ and the prepotential $F(X)$ of \cite{AFGNT}
are $2s$ and $2{\cal F}$ in this article. The matrix $\tilde{c}$ 
in \cite[(4.17)]{AFGNT} is also $2\Lambda_{\tilde{g}(X(v))}$ here. This 
systematic difference by a factor 2 is not a matter of convention;  
the normalization of ${\cal F}$ has been fixed unambiguously by 
(\ref{eq:4dSUSY-central-C}, \ref{eq:periods-Het-weak}) and/or  
(\ref{eq:gauge-kin-1}, \ref{eq:gauge-kin-2}), using an 
integral basis of charges. We demand that the matrices $\Lambda_{\tilde{g}}$
are integer valued. 
} %
\cite{AFGNT} 
\begin{align}
  M_{\tilde{g}(X(v))} = M({\bf 1}, \Lambda_{\tilde{g}(X(v))}),  \qquad 
  \Lambda_{\tilde{g}(X(v))} = - \frac{1}{2}
    \sum_{v' \in (\Q v \cap \widetilde{\Lambda}_S^\vee)}^{(-2\leq (v',v')^2<0)} n_{[v']} \;
      (v' \otimes v')_{IJ};  
     \label{eq:monodromy-matrix-gen-shiftpart}
\end{align}
the matrix $\Lambda_{\tilde{g}(X(v))}$ is $\Z$-valued, because 
contributions always come in pair from $v'$ and $-v'$, and 
$n_{[v']} = n_{[-v']}$.  

To determine all the monodromy representation matrices, 
the remaining task is to choose a set of generators $\{ g_i \}$ of 
$[\Gamma_S]$,  find a path $\tilde{\gamma}_{\tilde{g}_i}$ that starts from 
$(s_0, [\mho(t_0)])$ and ends at $(s_0^{\tilde{g}_i}, g_i \cdot [\mho(t_0)])$ 
for each element $g_i$, and to work out its monodromy matrices.  
All the paths in (\ref{eq:vect-moduli-sp-apprx-woX-noQnt}) from 
$(s_0, t_0)$ that become loops in the moduli 
space (\ref{eq:vect-moduli-sp-apprx-woX}) are obtained as 
compositions of those paths $\tilde{\gamma}_{\tilde{g}_i}$ along with the 
path $\tilde{\gamma}_D$ and the loops $\tilde{\gamma}_{\tilde{g}(X(v))}$. 
Such generators may be subject to some relations, but monodromy 
representation matrices should automatically satisfy the relations 
by construction. 

Here is the program we start to carry out in this article. 
We start off by assuming that there is a branch of moduli space 
whose classification invariants are 
$(\Lambda_S, \Lambda_T)$, $\{ \Phi_\gamma \}$, $\{ \Psi_\gamma \}$, 
$(a_{ab})_{+\Z}$ and $(b_a)_{+24\Z}$. 
By exploiting known facts, we choose a set of generators of $[\Gamma_S]$, 
find lifts $\tilde{\gamma}_{\tilde{g}}$ of those generators, and 
compute their monodromy matrices in terms of those invariants. When some 
of the monodromy matrices turn out not to be integer valued for 
those invariants, that is a contradiction. 
There should be no such branch of moduli space. This is how we obtain 
additional consistency conditions on the classification invariants. 

There are $\rho$ elements $g_{\infty(d)} \in \Gamma_S 
\subset [\Gamma_S]$ (where $d=1,\cdots, \rho$) 
whose monodromy matrices can be computed immediately. To be more 
precise, let $\tilde{\gamma}_{\tilde{g}_{\infty(d)}}$ be the path 
in (\ref{eq:vect-moduli-sp-apprx-woX-noQnt}) where $s$ remains constant 
at $s_0$, and $t^a$ varies from $t_0^a$ to $t_0^a + \delta^a_{\; d}$ in 
a straight line;
we choose $t_0$ so that ${\rm Im}(t^a_0) \gg 1$, and $t_0$ is far away 
from the loci $X(v)$ of extra 4d massless particles. Then 
\begin{align}
  (\Lambda_{\tilde{g}_{\infty(d)}})_{IJ} = \left( \begin{array}{cc|c}
    \left( \frac{d_{ddd}}{3}-a_{dd}-\frac{b_d}{12} \right) & 0 
         & \frac{d_{ddb}}{2} - a_{db} \\
    0 & 0 & 0 \\
   \hline 
    \left( \frac{d_{add}}{2} - a_{ad} \right) & 0 & d_{dab} 
  \end{array} \right)_{IJ}, 
 \label{eq:Lambda-infty-formula}
\end{align}
where $I\in \{0,\sharp, a\}$ and $J \in \{ 0,\sharp, b\}$. 
So, the following conditions are obtained:
\begin{align}
  d_{abd} \in \Z, \quad \frac{d_{dda}}{2}-a_{ad} \in \Z, \quad 
  \frac{d_{ddd}}{3} - a_{dd} - \frac{b_d}{12} \in \Z, \qquad \quad {}^\forall a,b,d \in \{1,\cdots, \rho\}. 
  \label{eq:cond-from-PQ-gen}
\end{align}
Without loss of information, the conditions (\ref{eq:cond-from-PQ-gen}) can 
also be stated as  
\begin{align}
 d_{dab} \in \Z, \quad  
  d_{add} + d_{aad} \in 2\Z, \qquad 2d_{ddd} + b_d \in 12 \Z 
  \label{eq:cond-from-PQ-gen-4Wall}
\end{align}
and
\begin{align}
  a_{ab} \in \frac{d_{abb}}{2} + \Z. 
  \label{eq:cond-from-PQ-gen-4Wall-open-a}
\end{align}

Two remarks are in order here. First, if we allow ourselves to 
replace the parameters $b_a$ mod $12\Z$ in the condition 
(\ref{eq:cond-from-PQ-gen-4Wall}) by $(c_2)_a$, then 
the condition (\ref{eq:cond-from-PQ-gen-4Wall}) is precisely equal 
to Wall's condition \cite{Wall} for existence of a diffeomorphism 
class of manifolds $[M]$ with the topological trilinear intersection 
(\ref{eq:def-geom-trilin}) and the second Chern class (\ref{eq:def-geom-c2})
given by $(C_{ab}, d_{abc})$ and $(24, (c_2)_a)$.  
When we start from a hypothetical branch of Het--Type IIA dual moduli space 
characterized by $(\Lambda_S, \Lambda_T)$, $\{ \Phi_\gamma \}$, 
$\{ \Psi_\gamma \}$, $(a_{ab})_{+\Z}$ and $(b_a)_{+24\Z}$, 
however, there is a priori no guarantee that $(c_2)_a$ mod 24 is equal 
to $(b_a)_{+24\Z}$. So, it is one of the tasks in this article 
to examine whether $(b_a)_{+24\Z}$ are equal to $(c_2)_a + 24\Z$; 
we wait until this check is done to conclude\footnote{
Alternatively, one may reason already at this moment 
(cf footnote \ref{fn:just-vect}) that a branch of moduli space 
satisfying (\ref{eq:cond-from-PQ-gen-4Wall}) must be realized by 
a Type IIA compactification on a manifold $M$, where $[M]$ is a 
diffeomorphism class whose existence is guaranteed by Wall's theorem; 
here, the data $(b_a)_{+24\Z}$ do not specify one diffeomorphism class 
uniquely due to the $+ 24\Lambda'_{\sharp a}$ ambiguity 
in (\ref{eq:abd-ambiguity}), but one may still think that 
one of those diffeomorphism classes will realize the branch in question. 
The remaining question then is to find out how the relation 
$(b_a)_{+24\Z} = (c_2)_a + 24\Z$ in geometric phases comes about 
theoretically; the study in the main text may therefore be regarded as 
addressing this question, besides the program of narrowing down 
the range of theoretically consistent classification invariants. 
} %
 that such a branch of moduli space has a region described by a non-linear 
sigma model of the target space $M$.  

Another thing to note at this moment is that only the parameters $d_{abc}$
are determined by the classification invariants $\{\Phi_\gamma \}$ and 
$\{\Psi_\gamma \}$, 
so only the two conditions out of (\ref{eq:cond-from-PQ-gen-4Wall}) set 
constraints on the integer parameters of $\{ \Phi_\gamma \}$ and 
$\{ \Psi_\gamma \}$.
Those integer parameters do not determine the real parts of $a_{ab}$ and 
$b_a$ as stated in (\ref{eq:matching-on-aNb}), so the 
condition (\ref{eq:cond-from-PQ-gen-4Wall-open-a}) and the last one 
in (\ref{eq:cond-from-PQ-gen-4Wall}) determine the values
of $(a_{ab})_{+ \Z}$ and of $(b_a)_{ + 24\Z} + 12\Z$. We will see in the 
case $\Lambda_S = \vev{+2}$ in section \ref{sec:deg2} that 
similar conditions from other generators of $[\Gamma_S]$ 
are combined with (\ref{eq:cond-from-PQ-gen-4Wall}) to yield 
further constraints on the integer parameters of $\{ \Phi_\gamma \}$ and 
$\{ \Psi_\gamma \}$. 

\subsection{Automorphic Forms for the Cases with 
${\rm rank}(\Lambda_S)=1$ and $\Lambda_S = U$}
\label{ssec:idea-autoMF}

To determine the matrices $\Lambda_{\tilde{g}}$ in 
$M_{\tilde{g}}(g, \Lambda_{\tilde{g}})$ for generators $\{ g \}$ of 
the lattice isometry group $[\Gamma_S]$, the following observation 
is useful. In the region with ${\rm Im}(s) \gg 1$ in the 
vector multiplet moduli space, various terms in the 
prepotential (\ref{eq:prepot-Het-weak}) can be grouped into 
\begin{align}
 {\cal F} = \frac{s}{2}(t,t) + {\cal F}^{(1)}(t) + {\cal O}(e^{2\pi i s}) 
\end{align}
by the $s$-dependence. The second term ${\cal F}^{(1)}$ is the 
1-loop correction in Heterotic string, which consists of the cubic 
polynomial in $t$ as well as $e^{2\pi i t}$ terms.
In the discussion around (4.26) of \cite{deWit}, 
the relation (\ref{eq:duality-map-proj-reltn}) is exploited to 
derive the relation
\begin{align}
 (g^{I=0}_{\; J} \mho(t)^J)^2 {\cal F}^{(1)}(t^{\tilde{g}}) = {\cal F}^{(1)}(t) 
    + \frac{1}{2} (\Lambda_{\tilde{g}})_{IJ} \mho(t)^I \mho(t)^J 
   \label{eq:deWit-1loopSuper-constraint}
\end{align}
for a duality transformation $\tilde{g} = M(g, \Lambda_{\tilde{g}})$. 
Readers are referred to \cite{deWit} for a detailed proof.\footnote{
Use $X^I F_I = (X^0)^2 (2{\cal F}^{(1)}+{\cal O}(e^{2\pi i s}))$ and 
$X^I = \mho^I(t) + {\cal O}(e^{2\pi i s})$. 
} %
We just note here that ${\cal F}^{(1)}(t^{\tilde{g}})$ on the left-hand side means 
the analytic continuation of ${\cal F}^{(1)}(t)$ along the path 
$\tilde{\gamma}_{\tilde{g}}$ in 
${\cal H}_s \times (D(\widetilde{\Lambda}_S) \backslash X_{\rm singl})$ 
from $t$ to $t^{g}$, seen as a function of the original coordinates $t$. 
The analytic continuation of the single component function 
${\cal F}^{(1)}$ along a path 
$\tilde{\gamma}_{\tilde{g}}$---its projection to $D(\widetilde{\Lambda}_S) 
\backslash X_{\rm singl}$ in fact---almost determines the matrix 
$\Lambda_{\tilde{g}}$ through (\ref{eq:deWit-1loopSuper-constraint}). 
The remaining ambiguity in the matrix $\Lambda_{\tilde{g}}$ [resp. $M_{\tilde{g}}$]
is $+n \widetilde{C}$ [resp. $\times (M_D)^{n}$] with $n \in \Z$, because 
$(\mho, \mho) = 0$. This ambiguity corresponds to the lost topological 
information when the path $\tilde{\gamma}_{\tilde{g}}$ is projected from 
$({\cal H}_s/\Z D) \times D(\widetilde{\Lambda}_S)$ to 
$D(\widetilde{\Lambda}_S)$. 

The easiest case where we can do this is for $\tilde{g}_{\infty(d)}$. 
\begin{align}
    {\cal F}^{(1)}(t^{\tilde{g}_{\infty(d)}}) - {\cal F}^{(1)}(t)   & \; = 
  \frac{d_{dbc}}{2}t^bt^c + \left(\frac{d_{ddb}}{2} - a_{db} \right) t^b 
  + \left( \frac{d_{ddd}}{6}-\frac{a_{dd}}{2} - \frac{b_d}{24}\right), 
  \label{eq:determine-Lambda-infty}
\end{align}
from which the matrix (\ref{eq:Lambda-infty-formula}) follows. 
We will see another example of easy analytic continuation in 
section \ref{sec:U}.

For some other generators $g$ of $[\Gamma_S]$, it is not always 
easy to work out the analytic continuation of ${\cal F}^{(1)}$ 
along a path $\tilde{\gamma}_{\tilde{g}}$. A combination of two  
observations in the literature comes to rescue, however, at least 
for some classes of lattices $\Lambda_S$. 

The first observation is that 
the function ${\cal F}^{(1)}$ on $D(\widetilde{\Lambda}_S)$ is almost 
an automorphic form of weight $-2$ under the lattice isometry group 
${\rm Isom}(\widetilde{\Lambda}_S)$; it has singularity along 
$X_{\rm singl}$, and is not precisely automorphic because of the 
second term in (\ref{eq:deWit-1loopSuper-constraint}). 
One may even hope that one of appropriate derivatives of ${\cal F}^{(1)}$ 
may turn into a genuine meromorphic automorphic form, because the second 
term in (\ref{eq:deWit-1loopSuper-constraint}) is an at most quartic 
polynomial in $t^a$'s. 

In the cases of ${\rm rank}(\Lambda_S)=1$ (where 
$\Lambda_S = \vev{+2k}$ with $k \in \{ 1,2,\cdots \} = \N$) 
and in the case $\Lambda_S = U$, 
the fifth derivative and the third derivative of ${\cal F}^{(1)}$, 
respectively, is indeed an automorphic form, denoted by $f_*$. 
The function ${\cal F}^{(1)}$ is an iterated integral of the 
automorphic form $f_*$ then. 
The violation of the automorphic transformation law of ${\cal F}^{(1)}$, 
the term involving the matrix $\Lambda_{\tilde{g}}$ 
in (\ref{eq:deWit-1loopSuper-constraint}), is determined by 
the property of $f_*$.  
More review is provided in sections \ref{sec:deg2} and \ref{sec:U}. 

The other observation is that those automorphic forms $f_*$ are 
determined by the classification invariants $\{ \Phi_\gamma \}$. 
An idea is that the function $G^{(1)}(t)$ in the 1-loop 
correction term of the K\"{a}hler potential of the vector-multiplet 
scalar fields, 
\begin{align}
 \hat{K}  & \; =: 
  - \ln\left[ \frac{s-\bar{s}}{i} \right] 
    - \ln \left[ - \frac{(t-\bar{t},t-\bar{t})}{2} \right]
   + \hat{K}^{(1)}(s,t) 
   + {\cal O}(({\rm Im}(s))^{-2}) + {\cal O}(e^{-2\pi {\rm Im}(s)}),   
\end{align}
\begin{align}
  \hat{K}^{(1)}(s,t) =: \frac{2i}{s-\bar{s}} G^{(1)}(t), 
\end{align}
is related both to\footnote{
\begin{align}
 G^{(1)}(t) = \frac{i}{(t-\bar{t},t-\bar{t})}
  \left[ 2(\overline{\cal F}^{(1)}-{\cal F}^{(1)})
  + (t-\bar{t})^d(\partial_d{\cal F}^{(1)} + \overline{\partial_d {\cal F}^{(1)}})
  \right] =: - \frac{V_{GS}}{4\pi},  
\end{align}
to be more explicit. So, $G^{(1)}_{a\bar{b}}$ can be expressed in terms of 
${\cal F}^{(1)}$. One can further 
derive (\ref{eq:f*-KG-relatn-rho=1}, \ref{eq:f*-KG-relatn-U}). 
} %
 ${\cal F}^{(1)}$ and also to $\{ \Phi_\gamma \}$. For a general $\Lambda_S$, 
the Laplacian of the potential $G^{(1)}$ is given in terms 
of the index $\{ \Phi_\gamma \}$ as 
follows \cite{AGNT-threshold,AFGNT,AP}:\footnote{
In the middle expression, the integral is over the $g=1$ worldsheet 
complex structure $\tau = \tau_{\rm Re} + i \tau_{\rm Im}$ in the 
Heterotic string language, within one fundamental region ${Fnd}$ of 
${\rm PSL}_2\Z$ in the upper half plane ${\cal H}$. 
In the integrand, $Z_\gamma := \sum_{v \in \gamma} q^{p_L(v)^2/2} \bar{q}^{p_R(v)^2/2}$, 
where $q = e^{2\pi i \tau}$, $\bar{q} = e^{-2\pi i \bar{\tau}}$, and 
$(p_L, p_R): \widetilde{\Lambda}_S^\vee \rightarrow \R^{\rho,2}$
is the momenta associated with the $(\rho,2)$ chiral bosons on 
the Heterotic string worldsheet. 
}\raisebox{5pt}{,}\footnote{
In the latter expression (cf \cite{DKL, HM}), 
\begin{align}
  \tilde{I}[F] & \; := \int_{Fnd} \frac{d\tau_{\rm Re}d\tau_{\rm Im}}{\tau_{\rm Im}}
   \sum_\gamma \left( Z_\gamma \hat{E}_2 F_\gamma - \tilde{c}_\gamma(0)\right),
  \qquad 
 I[F'] := \int_{Fnd} \frac{d\tau_{\rm Re}d\tau_{\rm Im}}{\tau_{\rm Im}} 
   \sum_\gamma \left( Z_\gamma F'_\gamma - c'_\gamma(0) \right) 
  \label{eq:def-HM-def-I}
\end{align}
for weight $-1-\rho/2$ and $1-\rho/2$ vector valued modular forms 
$F$ and $F'$, respectively; $\tilde{c}_\gamma(0)$ and $c'_\gamma(0)$ 
are Fourier coefficients in 
$F_\gamma E_2 =: \sum_\nu \tilde{c}_\gamma(\nu) q^\nu$ and  
$F'_\gamma(\tau) =: \sum_\nu c'_\gamma(\nu)q^\nu$, respectively, 
inserted so that the integrals $\tilde{I}[F]$ and $I[F']$ do not 
diverge.  When we use the Ramanujan--Serre derivative 
$\partial^S F_\gamma$ for $F'$, the constant $c'_0(0)$ 
for $F' = \partial^S F$ is equal to 
$[-(1+\rho/2)/12]$ times $\tilde{c}_0(0)$ for $F$, so those 
insertions do not have a net effect.  
}\raisebox{5pt}{,}\footnote{
\label{fn:KG-Phi-relatn}
The matrix $G^{(1)}_{a\bar{b}}$ is not always in a form of 
$\hat{K}^{(0)}_{a\bar{b}} {\cal I}$ for some scalar-valued integral 
function ${\cal I}$ of moduli $t$. For example, in the case of $\Lambda_S = U$, 
$G^{(1)}_{\rho\bar{u}} \neq 0$ but $\hat{K}^{(0)}_{\rho\bar{u}} = 0$. \\
We have confirmed by closely following the derivations 
in \cite{AGNT-threshold}, however, that the scalar-valued integral in 
(\ref{eq:KG-Phi-relatn}) determines the trace of the matrix $G^{(1)}_{a\bar{b}}$
for a general lattice $\Lambda_S$. 
} %
\begin{align}
 \hat{K}^{(0)a\bar{b}} G^{(1)}_{a\bar{b}} & \; = \frac{2i}{(4\pi)^2}
    \int_{Fnd}\frac{d\tau_{\rm Re} d\tau_{\rm Im}}{(\tau_{\rm Im})^{1+\frac{\rho}{2}}}
    \sum_\gamma F_\gamma \partial_\tau [(\tau_{\rm Im})^{\frac{\rho}{2}} Z_\gamma ] 
  = \frac{1}{32\pi} \left[ \frac{\rho +2}{3}I[F] - 8 I[\partial^SF]\right].
   \label{eq:KG-Phi-relatn}
\end{align}
The rest is to find a relation between $f_*$ and 
$\hat{K}^{(0)a\bar{b}}G^{(1)}_{a\bar{b}}$; the formulae quoted as 
(\ref{eq:f*-def-rho=1}, \ref{eq:f*-KG-relatn-rho=1}) 
and (\ref{eq:f*-def-U}, \ref{eq:f*-KG-relatn-U}) do that for the 
$\rho =1$ cases and the $\Lambda_S = U$ case, respectively.     

Now, it is not necessary to construct a Heterotic string 
$(c,\tilde{c}) = (22,9)$ SCFT, or a Type IIA $(c,\tilde{c}) = (9,9)$ SCFT. 
We may just assume that there exists a branch of moduli space with a set of 
classification invariants $(\Lambda_S, \Lambda_T)$ and $\{ \Phi_\gamma \}$, 
to get started. Compute the automorphic forms abstractly from the data 
$\Lambda_S$ and $\{ \Phi_\gamma \}$, and then determine the residual terms  
in the automorphic transformation of ${\cal F}^{(1)}$ 
in (\ref{eq:deWit-1loopSuper-constraint}), for generators $g$ 
of $[\Gamma_S]$. By demanding that the matrices $\Lambda_{\tilde{g}}$ 
should be $\Z$-valued, we will find out that string vacua cannot 
exist for certain choices of the classification invariants. 
That is what we do in the following sections. 

\section{The Case $\Lambda_S = \vev{+2}$}
\label{sec:deg2}

\subsection{Mathematical Facts Related to the Rank-1 Cases in General}
\label{ssec:rho=1-review}

To implement the general idea described in 
section \ref{ssec:idea-autoMF} for cases with various lattices 
$\Lambda_S$, there are more we can exploit from the literatures. 
In section \ref{ssec:rho=1-review}, we summarize those things 
for cases with $\rho = {\rm rank}(\Lambda_S)=1$. 

A rank-1 positive definite even lattice\footnote{
The rank-1 lattice $\Lambda_S = \vev{+2k}$ has a basis 
$\{e_{a=1} \}$ with the intersection form given by $(e_1, e_1) = 2k$. 
} %
has to be of the form 
$\vev{+2k}$ for $k=1,2,\cdots $. Any primitive embeddings  
$\Lambda_S=\vev{+2k} \hookrightarrow {\rm II}_{3,19}$ are identical 
modulo isometries of ${\rm II}_{3,19}$, and 
$\Lambda_T :=[(\Lambda_S)^\perp \subset {\rm II}_{3,19}]$ is 
determined modulo lattice isometry:
\begin{align}
    \widetilde{\Lambda}_S = U[-1] \oplus \vev{+2k}, \qquad 
   \Lambda_T = \vev{-2k} \oplus E_8^{\oplus 2} \oplus U^{\oplus 2}.
\end{align}
The following review is for a general $k \in \{ 1,2,\cdots \} =\Z_{>0}$,  
but we will pick up the $\Lambda_S = \vev{+2}$ case 
in section \ref{ssec:analysis-deg2} and carry out in detail 
the program in sections \ref{ssec:monodromy}--\ref{ssec:idea-autoMF}. 

\subsubsection{Lattice Isometry Group}

Here is what is known about the group of isometries of the lattice 
$\widetilde{\Lambda}_S$ (e.g., \cite{AP}). We begin 
with describing a group $\Gamma_0(k)_+$. Here, $k$ is a positive integer. 

The group $\Gamma_0(k)_+$ is a group of transformations on the complex 
upper half plane. It contains $\Gamma_0(k)/\{ \pm 1\}$ as a subgroup, 
where 
\begin{align}
  \Gamma_0(k) :=\left\{ \left. 
     g = \left(\begin{array}{cc} a & b \\ c & d \end{array}\right)
    \in {\rm SL}_2\Z \; \right| \; k|c \right\}, \qquad 
   g: {\cal H} \ni t \longmapsto \frac{at+b}{ct+d} \in {\cal H}.
\end{align}
Furthermore, there is an exact sequence 
\begin{align}
 1 \rightarrow \Gamma_0(k)/\{ \pm 1 \} \rightarrow \Gamma_0(k)_+ 
   \rightarrow (\Z_2)^s \rightarrow 1, 
  \label{eq:exct-seq-Gamma0k+}
\end{align}
where $s$ is the number of distinct primes appearing in the prime 
decomposition of $k$, $k= \prod_{i=1}^s p_i^{k_i}$. For an element 
of $(\Z_2)^s$, say $\epsilon = (\epsilon_1, \epsilon_2, \cdots, \epsilon_s)$
with $\epsilon_i \in \{ 0,1\}$, the coset $(\Gamma_0(k)/\{ \pm 1\}) 
\cdot \epsilon$ is described as follows. To prepare notations, let 
$k_1 := \prod_i p_i^{k_i \epsilon_i}$ and $k_2 := \prod_i p_i^{k_i (1-\epsilon_i)}$, 
so $k_1k_2 = k$ and $(k_1,k_2) =1$. The coset $(\Gamma_0(k)/\{ \pm 1\}) 
\cdot \epsilon$ is 
\begin{align}
  \left\{ \left. g= \pm 
  \left( \begin{array}{cc} k_1 a & b \\ k_1 c & d \end{array} \right) \right|
   \; a,b,c,d \in \Z, \; k_2|c \; k_1|d, \; (ad-bc)=1 \right\} ; 
  \quad g: t \longmapsto t^g := \frac{k_1a t + b}{k_1 c t+d}. 
  \label{eq:Gamma0k+-LFtransf}
\end{align}
Now, all the information is here to verify the exact sequence 
(\ref{eq:exct-seq-Gamma0k+}). 

The group $\Gamma_0(k)_+$ forms a subgroup of 
${\rm Isom}(\widetilde{\Lambda}_S)$. To see this, note that 
one can assign to an element $g$ of $\Gamma_0(k)_+$ 
in (\ref{eq:Gamma0k+-LFtransf}) the 
following linear transformation on $\widetilde{\Lambda}_S$:
for $n \in U[-1]\oplus \vev{+2k}$ described by 
$e_0 n^0+e_{\sharp} n^{\sharp} + e_{a=1} n^{a=1} =: 
(n^0, n^\sharp, n^1)^T \in U[-1] \oplus \vev{+2k}$, 
\begin{align}
 g: n \mapsto g\cdot n, \qquad 
 g = \left( \begin{array}{cc|c}
    d^2/k_1 & c^2/k_2 & 2 c d \\ k_2 b^2 &k_1  a^2 & 2k ab \\
    \hline bd/k_1 & ac/k_2 & ad+bc
 \end{array} \right). 
  \label{eq:Gamma0kP+Image}
\end{align}
This is a lattice isometry.\footnote{
That is, if $n\in \widetilde{\Lambda}_S$ then 
$g \cdot n\in \widetilde{\Lambda}_S$; furthermore,   
for any $n,n' \in \widetilde{\Lambda}_S$, 
$(g\cdot n', g \cdot n) = (n', n)$. 
} %
The assignment 
from $g \in \Gamma_0(k)_+$ to $g \in {\rm Isom}(\widetilde{\Lambda}_S)$
is an injective homomorphism, so our abuse of notation will be tolerated.  

Besides the image of $\Gamma_0(k)_+$, one will be aware of two more isometries; 
one is to multiply $(-1)$ to $\Lambda_S = \vev{+2k}$, denoted by $(-1)_{2k}$, 
and the other is to multiply $(-1)$ to $\widetilde{\Lambda}_S$, denoted 
by $-{\rm id}$. The group of all those isometries has a structure 
\begin{align}
  \left( \Gamma_0(k)_+ \rtimes \Z_2\vev{(-1)_{2k}} \right) \times 
   \Z_2\vev{-{\rm id}},  
\end{align}
and this is the group ${\rm Isom}(\widetilde{\Lambda}_S)$ in fact. 
Furthermore, any one of the isometries of $\widetilde{\Lambda}_S$ 
can be combined with an appropriate isometry on $\Lambda_T$ so that 
the pair of isometries defines an isometry of ${\rm II}_{4,20}$. 

The space $D(\widetilde{\Lambda}_S)$ is parametrized as 
\begin{align}
  \mho(t) = (1, kt^2, t)^T
\end{align}
with $t = e_{a=1} t^{a=1} \in \Lambda_S \otimes \C$ so that $(\mho, \mho)=0$; 
to be a little more precise, the space $D(\widetilde{\Lambda}_S)$ is 
$\{ t \in \C = \Lambda_S \otimes \C \; 
| \; {\rm Im}(t) \neq 0 \}$. The isometry $(-{\rm id})$ acts trivially 
on $D(\widetilde{\Lambda}_S)$, and $(-1)_{2k}: t \mapsto -t$ maps one 
connected component to the other. So,\footnote{
\label{fn:prd-dom-not-conn}
The moduli space $[\Gamma_S] \backslash D(\widetilde{\Lambda}_S)$ 
should actually be replaced by $[\Gamma_S] \backslash 
D(\widetilde{\Lambda}_S)/\Z_2$, where the $\Z_2$ action exchanges 
the role of $\mho$ and $\overline{\mho}$ (and $p_R^\C$ and 
$\overline{p_R^\C}$). So, we may restrict $D(\widetilde{\Lambda}_S)
 = \{ t \in \C \; | \; {\rm Im}(t) \neq 0\}$ to 
$\{ t \in \C \; | \; {\rm Im}(t) > 0\}$ to fix this $\Z_2$ gauge, 
when the $\Z_2 \subset {\rm Isom}(\widetilde{\Lambda}_S)$ gauge 
generated by $(-1)_{2k}$ is also fixed. 
} %
it is fine to focus on the subgroup $\Gamma_0(k)_+$ in studying 
the monodromy representations. 

\subsubsection{The Automorphic Form}

For $g \in \Gamma_0(k)_+$ of the 
form (\ref{eq:Gamma0kP+Image}, \ref{eq:Gamma0k+-LFtransf}), 
the relation (\ref{eq:deWit-1loopSuper-constraint}) reads   
\begin{align}
   \frac{(ck_1 t+d)^4}{k_1^2} {\cal F}^{(1)}(t^{\tilde{g}}) 
 = {\cal F}^{(1)}(t) + \frac{1}{2} (\Lambda_{\tilde{g}})_{IJ}\mho^I(t)\mho^J(t). 
   \label{eq:deWit-1loopSuper-constraint-rho=1}
\end{align}
After taking derivatives with respect to $t$ five times, the 2nd term 
on the right hand side drops.  On the left-hand side, 
one may use (\ref{eq:Gamma0k+-LFtransf}) and 
\begin{align}
  \frac{d t^g}{d t} = \frac{k_1}{(ck_1t+d)^2}, 
\end{align}
to arrive at \cite{KapLusThs, AP} (Bol's identity)
\begin{align}
  k_1^3 (\partial^5 {\cal F}^{(1)})(t^{\tilde{g}}) = (c k_1 t+ d)^6 \; 
   (\partial^5 {\cal F}^{(1)})(t). 
  \label{eq:rho1-F5-automorphic}
\end{align}
This means that 
\begin{align}
  f_*(t) := \frac{1}{(2\pi i)^5} (\partial_t^5{\cal F}^{(1)})(t)
  \label{eq:f*-def-rho=1}      
\end{align}
is a meromorphic automorphic form of weight-$(+6)$ of the group 
$\Gamma_0(k)_+$. 

The automorphic form $f_*$ is determined uniquely by the index 
$\{ \Phi_\gamma \}$. This is because, historically, 
there is a relation (\cite{AGNT-95-high-deriv} and \cite[(3.3)]{AP})
\begin{align}
 [t_2^{-4}\partial_t t_2^4][t_2^{-2} \partial_tt_2^2] \partial_t 
  \left( \hat{K}^{(0)t\bar{t}} G^{(1)}_{t\bar{t}} \right) = \cdots 
  = -\frac{i}{4k} (\partial_t^5 {\cal F}^{(1)}) 
   = - \frac{i}{4k}(2\pi i)^5 f_*(t). 
   \label{eq:f*-KG-relatn-rho=1}
\end{align}
So, one can just combine this with (\ref{eq:KG-Phi-relatn}). 
An alternative, and more practical perspective is to note that 
$f_* \propto \partial_t^5 {\cal F}^{(1)}$ contains only the 
coefficients $n_{w,0}$, which are all determined by $\{ \Phi_\gamma \}$ 
through (\ref{eq:matching-GV+chi}). 

The function ${\cal F}^{(1)}(t)$ is the five-fold iterated integral 
of the automorphic form $f_*(t)$ with respect to the coordinate 
$(2\pi i t)$. 
Although the integral ${\cal F}^{(1)}$ has a remnant of the 
automorphic transformation property of $f_*$, there is also 
a violation term in (\ref{eq:deWit-1loopSuper-constraint-rho=1}). 
Let us now turn to the question how the violation term is determined
from $f_*$. 

\subsubsection{Period Polynomials}
\label{sssec:PP}

Let $f(\sigma)$ be a cusp meromorphic modular form of weight-$w$ 
for a group $\Gamma$, and the group $\Gamma$ be either one of 
$\Gamma_0(k)_+$ for some $k$, or its finite index subgroup that 
at least contains the $\sigma \rightarrow \sigma + 1$ transformation. 
The automorphic form $f_*$ in (\ref{eq:f*-def-rho=1}) for 
the $\rho =1$ cases and $f_*$ in (\ref{eq:f*-def-U}) for the $\Lambda_S=U$ 
case satisfy this property; the weight is $(w=+6)$ and the argument 
$\sigma \in {\cal H}$ is $t$ in the $\rho=1$ cases, while $w=+4$ and 
$\sigma = u$ in the $\Lambda_S=U$ case.  
In the following, we will summarize necessary facts from the 
Eichler--Shimura theory of the modular transformation law of 
the $(w-1)$-fold iterated integral of $f(\sigma)$ \cite{AFGNT, AP}.
For more information, see e.g., \cite{Lang}. 

Let us begin with the following observation. 
The $(w-1)$-fold iterated integral along an arbitrary path
$\tilde{\gamma}$ starting from a base point $\sigma_0 \in {\cal H}$,  
\begin{align}
 I[\sigma; f,\tilde{\gamma}] := (2\pi i)^{w-1} \int_{\sigma_0}^{\sigma}d\sigma_1 \int_{\sigma_0}^{\sigma_1} d\sigma_2 \cdots \int_{\sigma_0}^{\sigma_{w-2}} d\sigma_{w-1} f(\sigma_{w-1}), 
\end{align}
has an alternative expression called the Eichler integral:
\begin{align}
  F_{\rm Eich}(\sigma;f, \tilde{\gamma}) :=  
  \frac{(2\pi i)^{w-1}}{(w-2)!} 
    \int_{\sigma_0}^\sigma d\sigma' \; f(\sigma') (\sigma-\sigma')^{w-2}. 
  \label{eq:2-exprsns-of-Eich-I}
\end{align}
Let us choose a base point $\sigma_0$ infinitesimally 
close to $i\infty$.
Note that the integral from $\sigma_0 \simeq i\infty$ is well-defined
in both $I[\sigma;f,\tilde{\gamma}]$ and $F_{\rm Eich}(\sigma;f,\tilde{\gamma})$
because of the cusp property of $f(\sigma)$. 

A general form of the $(w-1)$-fold indefinite integral is 
\begin{align}
 F(\sigma;f,\tilde{\gamma}) := I[\sigma;f,\tilde{\gamma}] + Q(\sigma) 
 = F_{\rm Eich}(\sigma;f, \tilde{\gamma}) + Q(\sigma), 
  \label{eq:add-intg-cnst-2EichlerI}
\end{align}
where $Q(\sigma)$ is a polynomial of $\sigma$ of degree at most $(w-2)$. 
The polynomial $Q(\sigma)$ cannot be determined from the modular form $f$,  
so it is a kind of integration constants.  

The Eichler integral has the following modular transformation property. 
For $g \in \Gamma_0(k)_+$ and a path $\tilde{\gamma}_{\tilde{g}}$ 
from $\sigma_0$ to $\sigma_0^g$, 
\begin{align}
  F_{\rm Eich}(\sigma^{\tilde{g}}; f, 
     (\tilde{\gamma})^g \circ \tilde{\gamma}_{\tilde{g}}) & \;
 =  \frac{(k_1 c\sigma + d)^{2-w}}{k_1^{1-w/2}}
     \left( F_{\rm Eich}(\sigma;f,\tilde{\gamma})
     + P_{\tilde{g}}(\sigma;f) \right), 
\end{align}
where 
\begin{align}
 P_{\tilde{g}}(\sigma, f) := \frac{(2\pi i)^{w-1}}{(w-2)!}
    \int_{(\tilde{\gamma}_{\tilde{g}})^{g^{-1}}} d\sigma \; f(\sigma') (\sigma - \sigma')^{w-2}.
  \label{eq:period-poly-def}
\end{align}
So, the $(w-1)$-fold iterated integral of a weight-$w$ modular form 
almost has a modular transformation property, with a weight 
$w + (w-1)(-2) = (2-w)$. How much the modular transformation property is 
violated is computed by the {\it period polynomials}
$P_{\tilde{g}}(\sigma, f)$, which is a polynomial of $\sigma$ of degree $(w-2)$. 

In most of math textbooks and literatures, Eichler integrals and 
period polynomials are introduced for modular forms that 
do not have a pole in the interior of the upper half plane. 
In the context of this article, however, we have no choice but 
to deal with modular forms $f$ that have a pole in the interior. 
So, we have to define Eichler integrals and period polynomials 
by specifying homotopy classes of integration contours that 
stay away from the poles of $f(\sigma)$. 

Let us now return to the original context in this article. 
Once the period polynomial is worked out for $\tilde{g}$, then
the polynomial (cf the appendix \ref{sec:Eichler-coh}) 
\begin{align}
   \frac{(k_1 c \sigma + d)^{w-2}}{k_1^{\frac{w}{2}-1}} 
          F(\sigma^g, f,(\tilde{\gamma})^g \circ \tilde{\gamma}_{\tilde{g}})
  - F(\sigma, f,\tilde{\gamma}) = 
  P_{\tilde{g}}(\sigma,f) +
     \frac{(k_1c\sigma+d)^{w-2}}{k_1^{\frac{w}{2}-1}} Q(\sigma^g)-Q(\sigma)
  \label{eq:PP+cobndry=anal.contn.violatn}
\end{align}
determines the matrix $\Lambda_{\tilde{g}}$ 
through (\ref{eq:deWit-1loopSuper-constraint-rho=1}) 
and (\ref{eq:deWit-1loopSuper-constraint-U}). 

For the group $\Gamma$, we need to choose a set of generators $\{ g_i \}$, 
their lifts $\tilde{g}_i$ and the paths $\tilde{\gamma}_{\tilde{g}_i}$ in 
$D(\widetilde{\Lambda}_S) \backslash X_{\rm singl}$. 
Those generators, in general, are not independent, but are subject 
to relations that follow from the composition law of the 
paths (\ref{eq:path-composition-law-2}) and homotopy equivalence of the paths. 
Once the monodromy representation matrices 
$M_{\tilde{g}_i}(g_i, \Lambda_{\tilde{g}_i})$ are found for those generators, 
however, those matrices automatically satisfy the relations. To see this, 
suppose that $\tilde{g}_a$, $\tilde{g}_b$ are some lifts of $g_a$ and 
$g_b$, respectively, and let $\tilde{g}_c := \tilde{g}_a \cdot \tilde{g}_b$ 
and $g_c := g_a \cdot g_b$, so $\tilde{\gamma}_{\tilde{g}_c} = 
(\gamma_{\tilde{g}_b})^{g_a} \circ \gamma_{\tilde{g}_a}$, as a reminder. 
Now, note that 
\begin{align}
  P_{\tilde{g}_c}(\sigma, f)  & \; = 
      \frac{\varphi_{g_b}(\sigma)^{w-2}}{k_{1,b}^{w/2-1}}
             P_{\tilde{g}_a}(\sigma^{g_b}, f)
   + P_{\tilde{g}_b}(\sigma, f),
  \label{eq:monodrm-product-compatible-PP}
\end{align}
and that 
\begin{align}
  \frac{\varphi_{g_c}(\sigma)^{w-2}}{(k_{1,b}k_{1,a})^{w/2-1}} & \; Q(\sigma^{g_c}) - Q(\sigma) 
  =  \label{eq:monodrm-product-compatible-cobndry} \\
& \left[ \frac{\varphi_{g_a}(\sigma^{g_b})^{w-2}}{k_{1,a}^{w/2-1}} Q((\sigma^{g_b})^{g_a})
     - Q(\sigma^{g_b})\right] \frac{\varphi_{g_b}(\sigma)^{w-2}}{k_{1,b}^{w/2-1}}
  + 
 \left[\frac{\varphi_{g_b}(\sigma)^{w-2}}{k_{1,b}^{w/2-1}} Q(\sigma^{g_b}) - Q(\sigma)\right],
  \nonumber 
\end{align}
where $\varphi_{g_b}(\sigma) = c_b k_{1,b} \sigma+ d_b$ using $c, k_1, d$ 
of $g=g_b$ in (\ref{eq:Gamma0k+-LFtransf}, \ref{eq:Gamma0kP+Image}) 
for $c_b$, $k_{1,b}$ and $d_b$.
Combining them together (with $f=f_*$), we automatically have 
\begin{align}
 \Lambda_{\tilde{g}_c} = g_b^T \cdot \Lambda_{\tilde{g}_a} \cdot g_b
    + \Lambda_{\tilde{g}_b}, \qquad 
   M_{\tilde{g}_a} \cdot M_{\tilde{g}_b} = M_{\tilde{g}_c}. 
  \label{eq:monodrm-product-compatible-matrix}
\end{align}
%

\subsection{Analysis on the $\Lambda_S = \vev{+2}$ Cases}
\label{ssec:analysis-deg2}

In this article, we pick up just one case $\Lambda_S = \vev{+2}$
from the series of $\rho = 1$ cases, and carry out the program 
outlined earlier. 

\subsubsection{A Quick Review}
\label{sssec:deg2-review}

{\bf Classification Invariants:} 

In the case of $\Lambda_S= \vev{+2}$, it is known
that $G_S = \Z_2$, the vector valued modular form $\{ \Phi_\gamma \}$
is parametrized by two free low-energy BPS indices $n_{0}, n_{1/2}$ that 
are allowed to take value in 
\begin{align}
  n_0 = -2, \qquad n_{1/2} \in \{ 0,1,2,3,4\}. 
\end{align}
Concrete expressions of $\{\Phi_\gamma \}$ for those $(n_0, n_{1/2})$ are 
found in \S \ref{ssec:formula-Phi} and references there; 
for notations and the range of $(n_0, n_{1/2})$, see \cite{EW19}. 
It is also known \cite{EW19} (cf also \cite{BW-16}) that 
one more classification 
invariant parameter is necessary besides the data $(n_0, n_{1/2})$, in order 
to distinguish known distinct branches of moduli space with the lattice 
$\Lambda_S=\vev{+2}$. It is 
\begin{align}
b_{\cal R} \in 2^{-1} \Z_{\geq 0}. 
  \label{eq:para-in-Psi-deg2}
\end{align}
For more information, see \cite{EW19}.\footnote{
It is not that one branch of moduli space has a unique value of $b_{\cal R}$. 
Reference \cite{EW19} assigned a value of $b_{\cal R}$ to a branch of moduli 
space by finding a branch of enhanced gauge symmetry (probe gauge group), 
and reading out 
the 1-loop beta function $b_{\cal R}$; 
when there are multiple branches of enhanced gauge symmetry, there are 
multiple values of $b_{\cal R}$ assigned to the 
original branch. It is the set of $b_{\cal R}$'s that is assigned to the 
original branch of moduli space, to be precise. The set of $b_{\cal R}$'s 
should be such that difference among those $b_{\cal R}$'s are in $6\Z$ so that 
those $b_{\cal R}$'s result in a consistent determination of the effective 
theory parameters in (\ref{eq:EW19-fit-deg2-4d}, \ref{eq:EW19-fit-deg2-4c2}). 
} %

Once one set of the classification invariants $n_0, n_{1/2}, b_{\cal R}$ is 
given, then \cite{EW19}\footnote{
Reference \cite{EW19} argued that $\delta n_a$ are integers 
with a language that is valid when the Type IIA description 
has a phase given by a Calabi--Yau-target non-linear sigma model
in the branch of enhanced gauge symmetry. In fact, we can argue 
that $\delta n_a \in \Z$ whether the enhanced symmetry branch has 
a phase of non-linear sigma model description or not; we can just apply 
the argument leading to (\ref{eq:cond-from-PQ-gen}, \ref{eq:cond-from-PQ-gen-4Wall}, \ref{eq:cond-from-PQ-gen-4Wall-open-a}) to 
the branch of enhanced gauge symmetry. 
} %
\begin{align}
  d_{111} & \; = 4-b_{\cal R} - n_{1/2} + 6 \delta n_{a=1},
    \label{eq:EW19-fit-deg2-4d}   \\
  (c_2)_1 & \; = 52 - 4 b_{\cal R} - 10 n_{1/2} + 24 \delta n_{a=1}; 
    \label{eq:EW19-fit-deg2-4c2}
\end{align}
here, $d_{111}$ is the effective theory parameter $d_{abc}$ 
in (\ref{eq:prepot-Het-weak}), and $(c_2)_1$ is a coefficient 
appearing in the holomorphic $R^2$ term in the 4d effective 
theory (mentioned briefly at (\ref{eq:def-geom-c2})). 
The two effective theory parameters $d_{111}$ and $(c_2)_1$ are determined 
modulo $\delta n_1 \in \Z$; this ambiguity corresponds to the symplectic 
transformations $M({\bf 1}, \Lambda)$ with $\Lambda_{\sharp 1} 
= - \Lambda_{01} = \delta n_1 \in \Z$, changing the flat coordinate $s$ 
by $\Z t^{a=1}$.

We will see by (\ref{eq:deg2-b-div-by-2}), however, that not all 
$b_{\cal R} \in 2^{-1} \Z$ are theoretically possible. 

{\bf Isometry Group and Loci of Extra Massless Fields}

The group $\Gamma_0(k)_+ \subset {\rm Isom}(\widetilde{\Lambda}_S)$ is 
now ${\rm PSL}_2\Z$. Because ${\rm Isom}(G_S,q_S)$ is trivial for 
$\Lambda_S = \vev{+2k}$, all the isometries of $\widetilde{\Lambda}_S$ 
lift to isometries of the lattice ${\rm II}_{4,20}$. So, 
we demand that all the elements $g$ of ${\rm PSL}_2\Z$ have corresponding 
duality transformations $\tilde{g}$ and matrices $M_{\tilde{g}}$ in 
${\rm Sp}(2(\rho+2);\Z)^{\rm H.el}$. 

As argued already, it is enough to find a set of generators $\{ g_i \}$ 
of ${\rm PSL}_2\Z \subset [\Gamma_S]$, and construct their lifts, 
$\tilde{\gamma}_{\tilde{g}_i}$ and $M_{\tilde{g}_i}$; the matrices 
$M_{\tilde{g}_i}$ automatically satisfy appropriate relations that follow 
from the relations of $\{ g_i \}$'s and $\tilde{\gamma}_{\tilde{g}_i}$'s. 
We choose $\{ g_{\infty}^{\pm 1}, g_2^{\pm 1}\}$ as a set of generators; 
$g_\infty : D(\widetilde{\Lambda}_S)  \ni t \mapsto t^{g_\infty} = 
t+1 \in D(\widetilde{\Lambda}_S)$, and $g_2: t \mapsto t^{g_2} = -1/t$. 
The map $g_3: t \mapsto t^{g_3} = -1/(t+1)$ is obtained as 
$g_3 = g_2^{-1} \cdot g_\infty = g_2 \cdot g_\infty$. 

The loci of extra massless fields $X_{\rm singl}$ consist of the 
${\rm PSL}_2\Z$ orbits of 
\begin{align}
  X(v^{[0]}_*) & \; = \{ t = t^{[0]}_* := i \}, \qquad v^{[0]}_* = (1,1,0), \\
  X(v^{[1]}_*) & \; = \{ t= t^{[1]}_* := e^{2\pi i/3} \}, \qquad 
   v^{[1]}_* = (1,1,1),  
\end{align}
where we have used an integral basis $(e^0, e^\sharp, e^{a=1})^T$ of 
$\widetilde{\Lambda}_S^\vee$ to 
express $v^{[0]}_*$ and $v^{[1]}_*$ in $\widetilde{\Lambda}_S^\vee$. 
For a loop $\tilde{\gamma}$ in $D(\widetilde{\Lambda}_S)$ that goes 
around $X(v^{[0]}_*)$ [resp. $X(v^{[1]}_*)$] by phase $+2\pi$, the 
monodromy matrix is given by (\ref{eq:monodromy-matrix-gen-shiftpart}), 
with the data $n_{0}$ [resp. $n_{1/2}$] determining the matrices 
$\Lambda_{\tilde{g}(X(v^{[0]}_*))}$ [resp. $\Lambda_{\tilde{g}(X(v^{[1]}_*))}$]. 

We choose the base point $t_0$ in $D(\widetilde{\Lambda}_S)$ infinitesimally 
close to $+i\infty$. To choose a lift duality transformation 
$\tilde{g}_i$ for $g_i \in [\Gamma_S]$, a path $\tilde{\gamma}_{\tilde{g}_i}$ 
from $t_0$ is specified as follows: the path 
$\tilde{\gamma}_{\tilde{g}_\infty}$ is a straight line from $t_0$ to $t_0+1$ 
in the large ${\rm Im}(t)$ region of $D(\widetilde{\Lambda}_S)$; the 
path $\tilde{\gamma}_{\tilde{g}_2}$ is a path from $t_0 \simeq +i\infty$
to $t_0^{g_2} \simeq +i\epsilon$ almost straight down the imaginary axis 
in the complex $t$-plane that avoids $t=t^{[0]}_* = +i$ by detouring into the 
2nd quadrant (see Figure \ref{fig:cntr-deg2-forPP} (a)). We choose the path 
$\tilde{\gamma}_{(\tilde{g}_2)^{-1}}$ almost the same as 
$\tilde{\gamma}_{\tilde{g}_2}$, but it detours around the point $t=t^{[0]}_*=+i$ 
by stepping into the 1st quadrant of the $t$-plane. 
One can verify by using the path composition 
rule (\ref{eq:path-composition-law-2}) that $\tilde{g}_2^{-1}$ introduced 
in this way is indeed the inverse element of $\tilde{g}_2$.

{\bf The monodromy matrix $M_{\tilde{g}}$ for $\tilde{g} = \tilde{g}_\infty$:}

We have already discussed 
in (\ref{eq:Lambda-infty-formula}--\ref{eq:cond-from-PQ-gen-4Wall}) 
what the monodromy matrix $M_{\tilde{g}_\infty}$ should be. In the 
present context ($\rho=1$), the matrix $M_{\tilde{g}_\infty}$ is 
integer valued if and only if   
\begin{align}
 d_{111} \in \Z, \qquad a_{11} \in \frac{d_{111}}{2} + \Z, \qquad 
   2d_{111} + b_1 \in 12\Z.  
  \label{eq:cond-from-PQ-deg2-B}
\end{align}
It follows immediately from (\ref{eq:cond-from-PQ-deg2-B}) that 
$a_{11} \in \Z/2$ and $b_1 \in \Z$. Furthermore, the classification 
invariants should be subject to 
\begin{align}
  b_{\cal R} \in - n_{1/2} + \Z, 
\end{align}
so we should have the parameter $b_{\cal R}$ in $\Z$ not in $2^{-1} \Z$. 
The condition (\ref{eq:cond-from-PQ-deg2-B}) determines $(b_1)_{+24\Z}$ 
only mod $+12\Z$, so there are still two possible values of $(b_1)_{+24\Z}$. 
To compare $(b_1)_{+24\Z}$ and $(c_2)_1$ modulo $+12\Z$, 
\begin{align}
  (c_2)_1 \equiv 4 -4b_{\cal R} + 2n_{1/2}, \qquad 
  (b_1)_{+24\Z} \equiv (-2d_{111}) \equiv -8 + 2b_{\cal R} + 2n_{1/2}. 
  \label{eq:deg2-compare-b1-n-c21-mod12}
\end{align}
So, $(c_2)_1$ and $(b_1)_{+24\Z}$ are different when compared 
mod $12\Z$, if $b_{\cal R}$ is odd. They are equal mod $12\Z$ 
when $b_{\cal R}$ is even.

\subsubsection{Monodromy Matrix of $\tilde{g}_2$: the case $n_{1/2}=0$}

Let us now determine the monodromy matrix $M_{\tilde{g}_2}$ for 
the other generator of ${\rm PSL}_2\Z$. To find out the matrix 
$\Lambda_{\tilde{g}_2}$ by analytic continuation, we use the method 
described in sections \ref{ssec:idea-autoMF} and \ref{ssec:rho=1-review}.  
It is enough to evaluate the period polynomials 
in (\ref{eq:PP+cobndry=anal.contn.violatn}). 

To start, let us work on the case $n_{1/2}=0$. The automorphic form 
$f_*$ in (\ref{eq:f*-def-rho=1}) is of weight $(w=6)$, under the group 
${\rm PSL}_2\Z$. It is determined uniquely by the combination 
of (\ref{eq:f*-KG-relatn-rho=1}) and (\ref{eq:KG-Phi-relatn})
in terms of the indices $\{ \Phi_\gamma \}_{(n_0,n_{1/2})=(-2,0)}$.   
It is known that \cite{KapLusThs}
\begin{align}
  f_*(t) & \; := \frac{1}{(2\pi i)^3} \frac{(18E_4^3-5E_6^2)(E_4^3-E_6^2)}{9E_6^3}
      \label{eq:f*-def-n1=0} \\
  & \;  \simeq  \frac{1}{(2\pi i)^3} \left( 2496 \; {\rm Li}_{-2}(q)
    + 2^5 \cdot 223752 \; {\rm Li}_{-2}(q^2) + \cdots \right) 
\end{align}
is the right choice. One way to argue for (\ref{eq:f*-def-n1=0}) is 
to note that $f_*$ satisfies all the properties expected from physics, 
including appropriate singularity\footnote{
The Laurent series expansion of $f_*(t)$ in (\ref{eq:f*-def-n1=0}) is 
\begin{align}
 f_*(t) \simeq \frac{1}{(2\pi i)^6} \frac{-16}{(t-i)^3} + {\cal O}((t-i)^{-2}), 
\end{align}
from which $F_0 \simeq 8i(t-i) \ln[t-i]/(2\pi i) + {\cal O}((t-i)^2)$ 
and 
$F_1 \simeq -8(t-i) \ln[t-i]/(2\pi i)+{\cal O}((t-i)^2)$ follow. 
} %
 at $t=t^{[0]}_* = i$
and non-singular behavior at $t=t^{[1]}_* = e^{2\pi i/3}$. 
A more practical way is to see that the coefficients $2496$, $223752$, etc. 
of $w^5 {\rm Li}_{-2}(q^w)$ (for $w = e^{a=1} w \in \Lambda_S^\vee$) agree 
with $n_{w,0} = c_{[w]}((w,w)/2)$ (see \S \ref{ssec:formula-Phi}).

\begin{figure}[tbp]
  \begin{center}
   \begin{tabular}{ccccc}
    \includegraphics[width=0.2\linewidth]{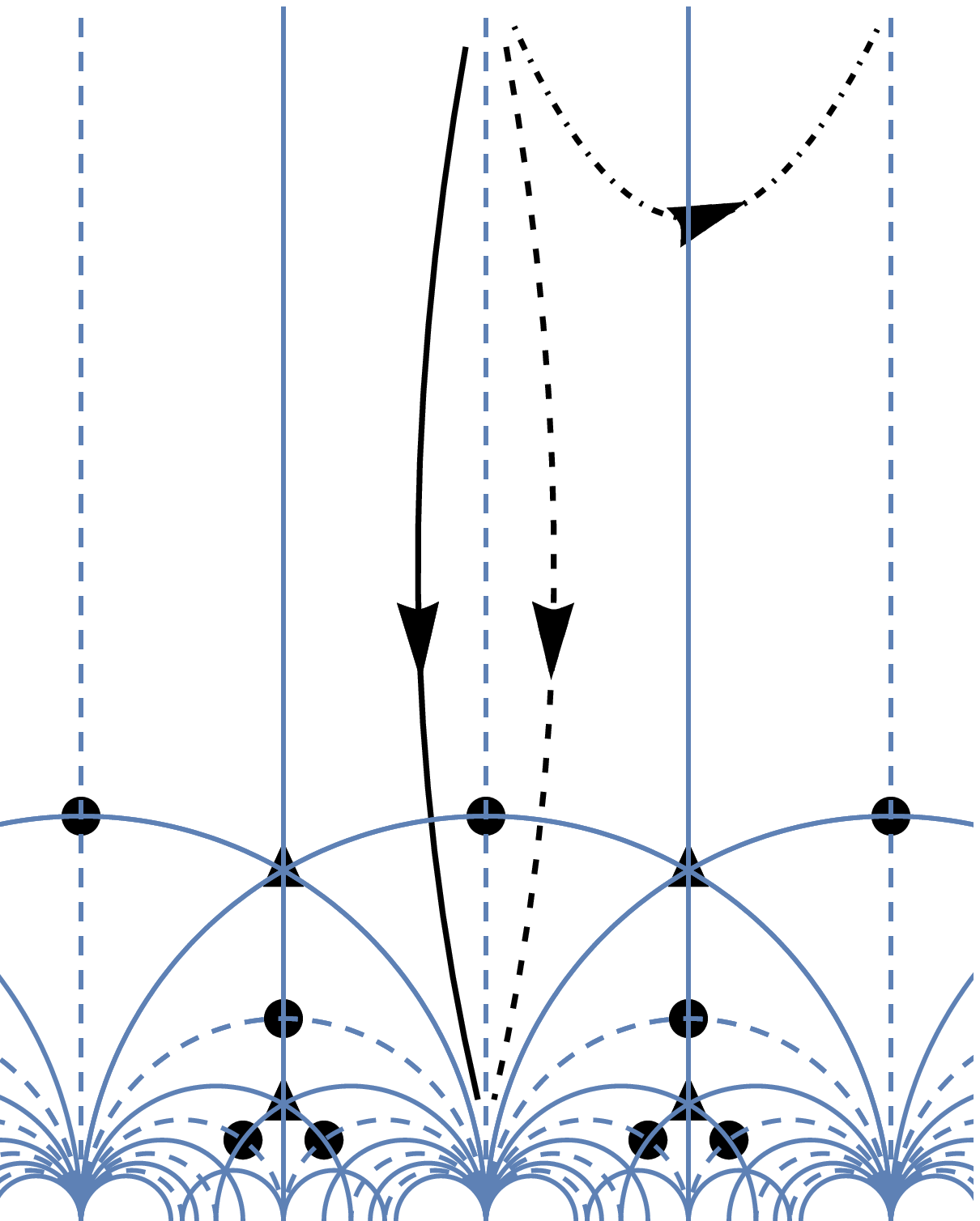} & $\quad$ &
    \includegraphics[width=0.2\linewidth]{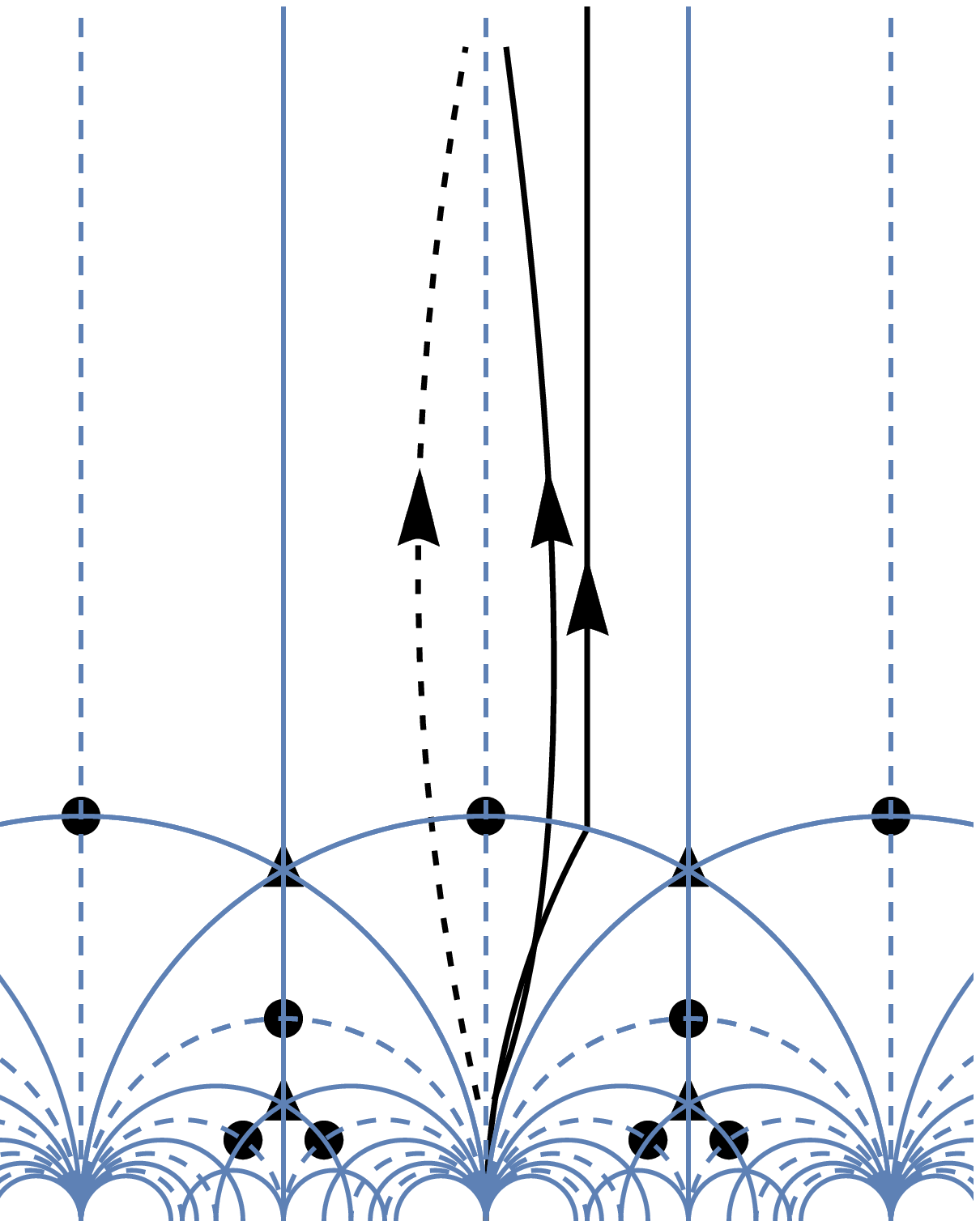} & $\quad$ &
    \includegraphics[width=0.2\linewidth]{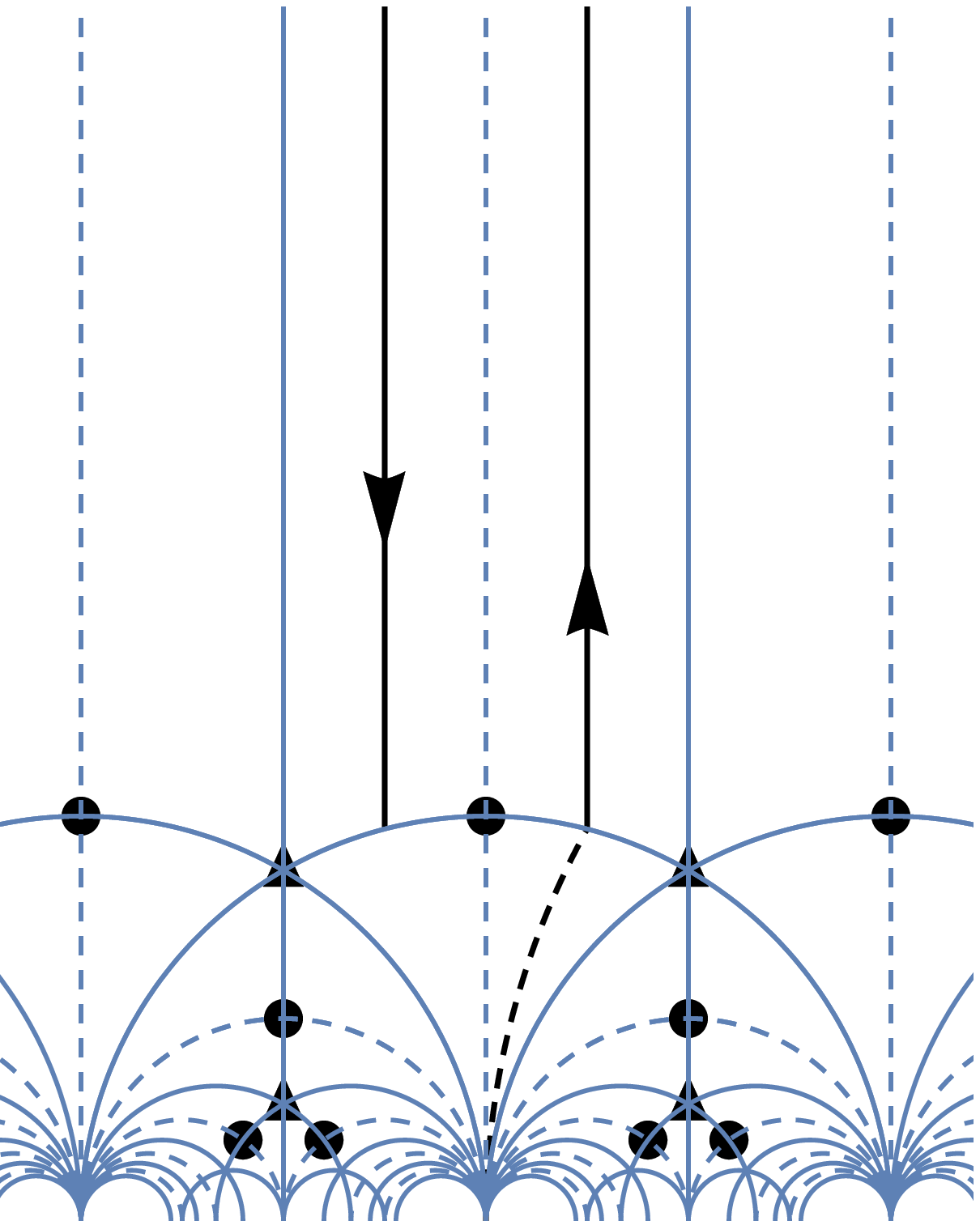} \\ 
  (a) & & (b) & & (c) 
   \end{tabular}
   \caption{\label{fig:cntr-deg2-forPP}
Paths $\tilde{\gamma}_{\tilde{g}_\infty}$, $\tilde{\gamma}_{\tilde{g}_2}$ and 
$\tilde{\gamma}_{\tilde{g}_2^{-1}}$ in the upper half plane $t \in {\cal H}$ 
are shown by dot-dashed, solid and dashed lines in (a) along with 
the boundaries of fundamental regions in thinner solid lines.   
The ${\rm PSL}_2\Z$ orbit of $t^{[0]}_*$ and $t^{[1]}_*$ are indicated 
by $\bullet$'s and $\blacktriangle$'s, respectively. 
The contours of integration for the period polynomials 
$P_{\tilde{g}_2}(t,f_*)$ and $P_{\tilde{g}_2^{-1}}(t,f_*)$ are drawn by solid and 
dashed lines, respectively, in (b). The contour for $P_{\tilde{g}_2}(t,f_*)$ may 
be split into two segments (solid and dashed) in (c), and the integral over 
the dashed part of the contour may be further rewritten as an integral 
over the other solid line contour in (c).  }
  \end{center}
\end{figure}

Now, we wish to compute the period polynomials $P_{(\tilde{g}_2)^{\pm 1}}(t,f_*)$. 
As we have chosen $t_0 \simeq +i\infty$ and the path 
$\tilde{\gamma}_{\tilde{g}_2}$ [resp. $\tilde{\gamma}_{\tilde{g}_2^{-1}}$] 
to be in the 2nd [resp. 1st] quadrant of the $t$-plane, 
the integration contour $(\tilde{\gamma}_{\tilde{g}_2})^{g_2^{-1}}$ [resp. 
$(\tilde{\gamma}_{\tilde{g}_2^{-1}})^{g_2}$] is 
in the 1st [resp. 2nd] quadrant, from $+i\epsilon$ to $+i\infty$ 
(Figure \ref{fig:cntr-deg2-forPP} (b)). 
The numerical integration of (\ref{eq:period-poly-def}) along this contour 
can be split into two segments by exploiting the modular transformation 
property of $f_*$ and change of variables 
(Figure \ref{fig:cntr-deg2-forPP} (c)). 
It is not hard to find out numerically by using Mathematica that 
\begin{align}
 P_{(\tilde{g}_2)^{\pm 1}}(t, f_*) & \; \simeq  - 0.610599 i (t^4-1)
   \pm 0.5(t^4+2t^2+1) -1.5(t^3+t),  \\
  & \; \simeq \frac{\zeta(3)}{(2\pi i)^3}\frac{(-252)}{2}(t^4-1)
    \pm \frac{t^4+2t+1}{2} - \frac{3}{2}(t^3+t).
  \label{eq:PP-deg2-n12=0-result}
\end{align}

Now, let us add the integration constants in the $(w-1)=5$-fold iterated 
integral, as in (\ref{eq:add-intg-cnst-2EichlerI}). The polynomial $Q(t)$
should be at most degree-4, but we set it to   
\begin{align}
Q(t) = \frac{ 0}{4!} t^4 + \frac{d_{111}}{3!} t^3
    - \frac{a_{11}}{2}t^2 - \frac{b_1}{24} t
    - \frac{\zeta(3)}{(2\pi i)^3}\frac{\chi}{2},  
  \label{eq:Qt-rho1-b}
\end{align}
because we know for a Het--IIA dual vacuum that there must be a 
symplectic frame where the quartic term is absent in the prepotential 
(\ref{eq:prepot-Het-weak}). 
The relation (\ref{eq:deWit-1loopSuper-constraint-rho=1}, 
\ref{eq:PP+cobndry=anal.contn.violatn}) determines the matrix 
$\Lambda_{\tilde{g}_2^{\pm 1}}$ as follows:  
\begin{align}
  \Lambda_{(\tilde{g}_{2})^{\pm 1}} = \left( \begin{array}{cc|c}
     \pm 1+\frac{\zeta(3)}{(2\pi i)^3}(\chi+252) & \pm 1
           & \left( \frac{b_1}{24}-\frac{d_{111}}{6}-\frac{3}{2}\right) \\
     \pm 1 & \pm 1-\frac{\zeta(3)}{(2\pi i)^3}(\chi+252)
           & \left( \frac{b_1}{24}-\frac{d_{111}}{6}-\frac{3}{2}\right) \\
   \hline
    \left( \frac{b_1}{24}-\frac{d_{111}}{6}-\frac{3}{2}\right) &
    \left( \frac{b_1}{24}-\frac{d_{111}}{6}-\frac{3}{2}\right) & 0 
  \end{array} \right) + \C \widetilde{C} . 
  \label{eq:deg2-Lg2-result}
\end{align}
The $+\C \widetilde{C}$ ambiguity should be reduced to $+\Z \widetilde{C}$ 
(or dropped) because the ambiguity beyond $+\Z \widetilde{C}$ 
does not help making this $\Lambda_{\tilde{g}_2}$ matrix integer valued. 

Before proceeding further, let us run a few checks with known results, 
to validate this method of computing the monodromy matrices. 
There is a branch with $\Lambda_S = \vev{+2}$ that has been 
studied extensively in the literature. That is the Heterotic construction 
known as the $ST$-model, whose Type IIA dual is for a Calabi--Yau 
threefold $M = (12) \subset \P^4_{[1:1:2:2:6]}$. This geometry indicates 
that $d_{111}=4$, $b_{1}=52$, and $\chi=-252$. When those values are 
substituted into (\ref{eq:deg2-Lg2-result}), the monodromy matrix 
$M_{\tilde{g}_2^{-1}} = M(g_2^{-1}, \Lambda_{\tilde{g}_2^{-1}})$ reproduces 
the monodromy matrix $(-S_1)$ in \cite{Non-P}, which was determined 
by using computations of the mirror manifold of $M$. 

As a second check, remember that we have chosen the two paths 
$\tilde{\gamma}_{\tilde{g}_2}$ and $\tilde{\gamma}_{\tilde{g}_2^{-1}}$ 
both as lifts of 
\begin{align}
 g_2  = \left( \begin{array}{cc|c}
    & 1 & \\ 1 & & \\ \hline & & -1 \end{array} \right)
   \in {\rm Isom}(\widetilde{\Lambda}_S) 
  \label{eq:deg2-matrix-g2}
\end{align}
in a way they are for the inverse duality transformation 
of each other under 
the path composition law (\ref{eq:path-composition-law-2}). 
Indeed, one can verify that both $M(g_2, \Lambda_{\tilde{g}_2}) \cdot 
M(g_2, \Lambda_{\tilde{g}_2^{-1}})$ and $M(g_2, \Lambda_{\tilde{g}_2^{-1}}) \cdot 
M(g_2, \Lambda_{\tilde{g}_2})$ are equal to the identity matrix 
by using (\ref{eq:deg2-Lg2-result}, \ref{eq:deg2-matrix-g2}); we 
do not need to use a specific value for $\chi$, $d_{111}$ and $b_1$. 
One will also find that 
\begin{align}
 \left( M(g_2,\Lambda_{\tilde{g}_2}) \right)^2
      = M({\bf 1}, \Lambda_{\tilde{g}(X(v^{[0]}_*))}), \qquad 
 \left( M(g_2, \Lambda_{\tilde{g}_2^{-1}}) \right)^2 
      = M({\bf 1}, - \Lambda_{\tilde{g}(X(v^{[0]}_*))}), 
\label{eq:deg2-monodromy-check-2}
\end{align}
without substituting any value into $\chi$, $d_{111}$ and $b_1$. 

\begin{figure}[tbp]
 \begin{center}
  \begin{tabular}{ccc}
    \includegraphics[width=0.30\linewidth]{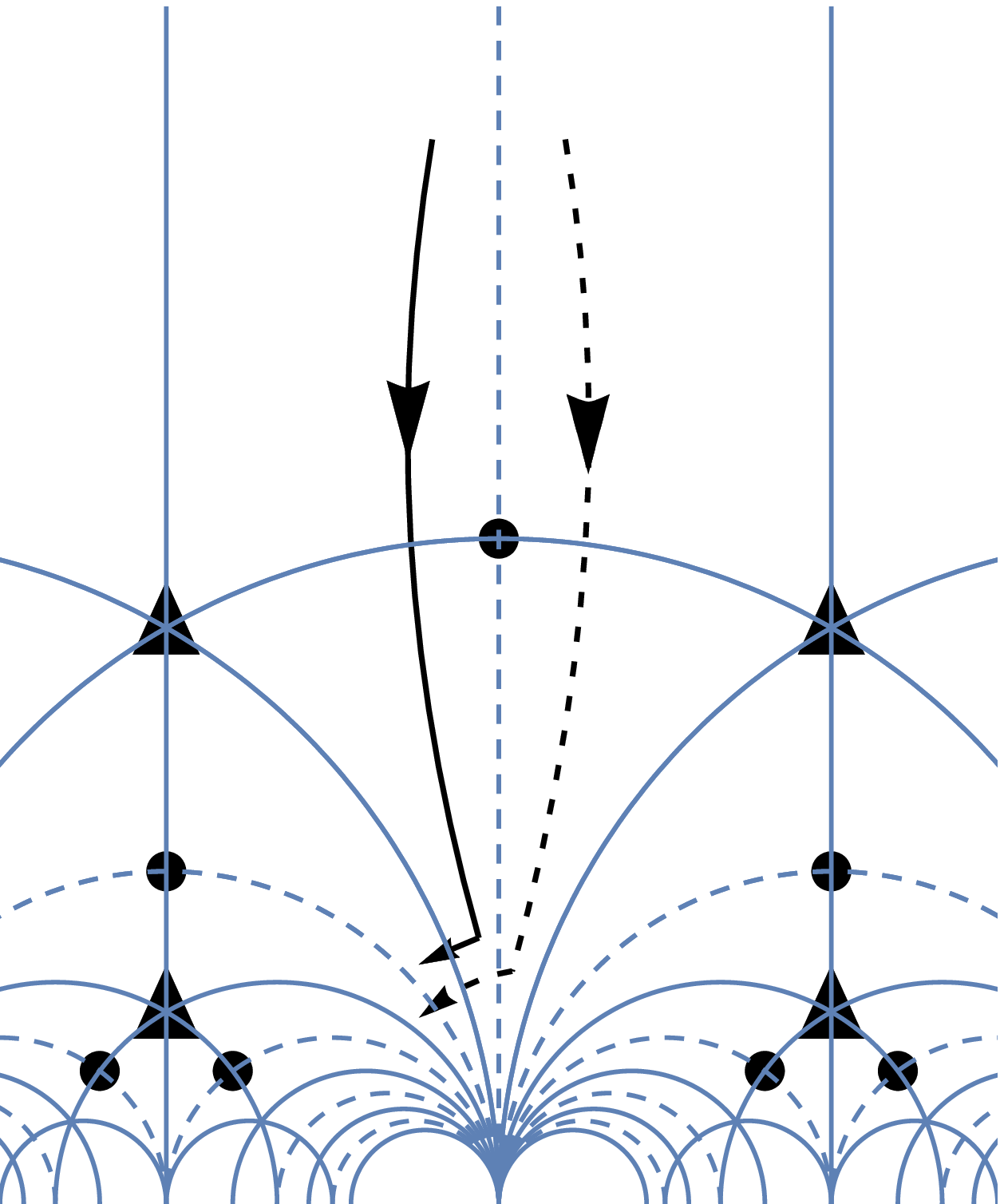} & $\quad$ &
    \includegraphics[width=0.44\linewidth]{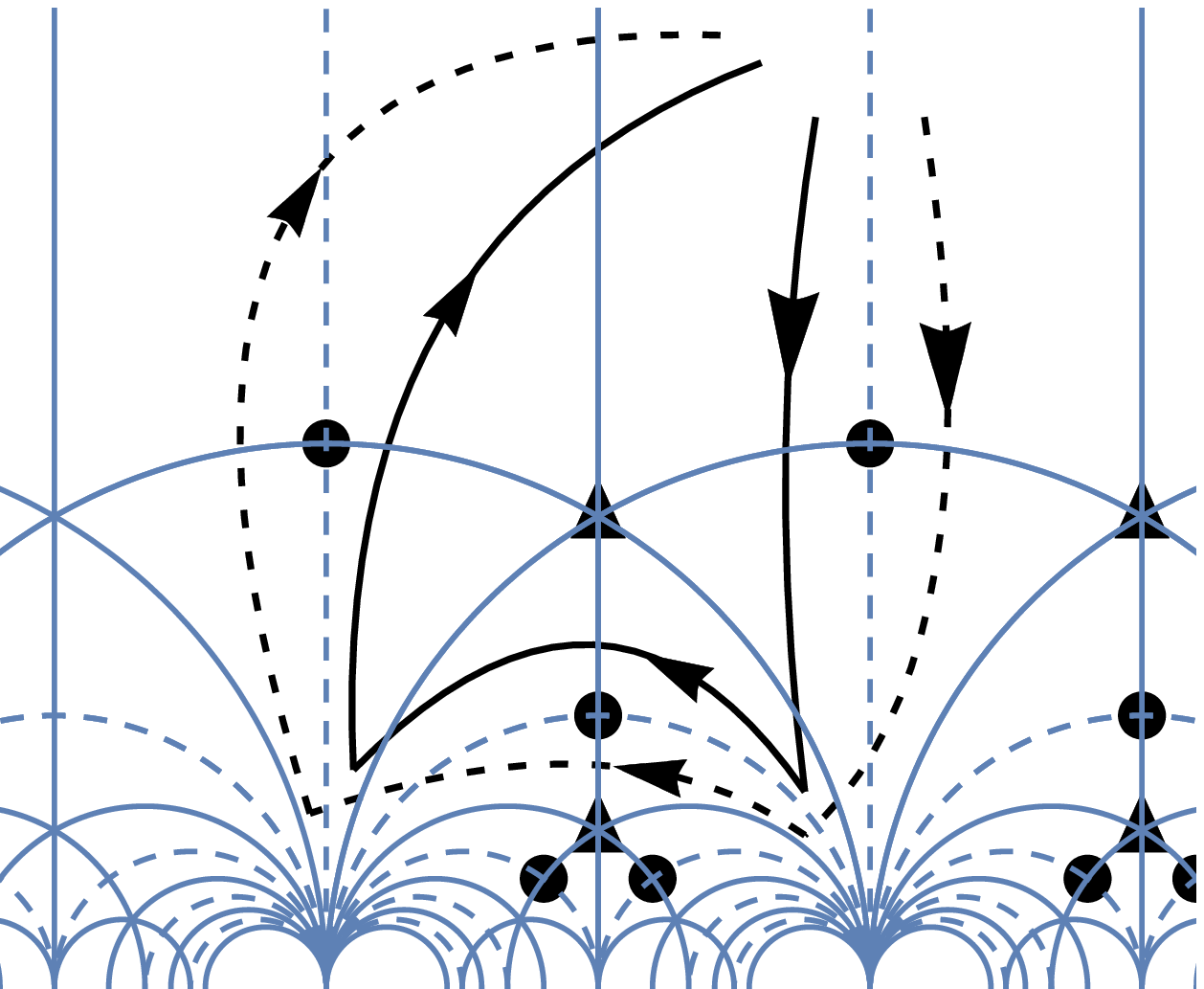} \\
    (a) & & (b) 
  \end{tabular}
  \caption{\label{fig:cntr-gamma3-deg2} The paths 
$\tilde{\gamma}_{\tilde{g}_3} = (\tilde{\gamma}_{\tilde{g}_\infty})^{g_2} \circ
 \tilde{\gamma}_{\tilde{g}_2}$ for 
$\tilde{g}_3 = \tilde{g}_2 \cdot \tilde{g}_\infty$ and 
$\tilde{\gamma}_{\tilde{g}'_3} = (\tilde{\gamma}_{\tilde{g}_\infty})^{g_2} \circ
\tilde{\gamma}_{\tilde{g}_2^{-1}}$ for $\tilde{g}'_3 =
 \tilde{g}_2^{-1} \cdot \tilde{g}_\infty$ are drawn in the complex $t$ 
plane in (a) by the solid and dashed oriented lines, respectively. 
In the panel (b), the solid and dashed oriented lines are the paths 
(loops in fact) for $(\tilde{g}_3)^3$ and $(\tilde{g}'_3)^3$, respectively. 
For more information, 
see the caption of Figure \ref{fig:cntr-deg2-forPP}. 
 }
 \end{center}
\end{figure}

Finally, one may define two lifts of the map $g_3: t \mapsto -1/(t+1)$. 
One is $\tilde{g}_3 := \tilde{g}_2 \cdot \tilde{g}_\infty$ and the 
other $\tilde{g}_3' := \tilde{g}_2^{-1} \cdot \tilde{g}_\infty$; note 
that $g_2 \cdot g_3 = g_\infty$ and $g_2^2 = 1$ in 
${\rm Isom}(\widetilde{\Lambda}_S)$. The corresponding paths 
$\tilde{\gamma}_{\tilde{g}_3}$ and $\tilde{\gamma}_{\tilde{g}'_3}$ 
are determined from (\ref{eq:path-composition-law-2}) and are 
shown in Figure \ref{fig:cntr-gamma3-deg2}~(a). 
A straightforward matrix computation 
confirms that 
\begin{align}
 (M_{\tilde{g}_3})^3 & \; = (M_{\tilde{g}_2}\cdot M_{\tilde{g}_\infty})^3 
 = M({\bf 1}, (d_{111}/2-2) \widetilde{C}), \label{eq:deg2-monodromy-check-3A} \\
 (M_{\tilde{g}'_3})^3 & \; = (M_{\tilde{g}_2^{-1}} \cdot M_{\tilde{g}_\infty})^3 
 = M({\bf 1}, \Lambda'), \\
 & \Lambda' = - \Lambda_{\tilde{g}(X(v^{[0]}_*))} - \Lambda_{\tilde{g}(X(v^{[0]}_{**}))}
  - \Lambda_{\tilde{g}(X(v^{[0]}_{***}))} + (d_{111}/2-2) \widetilde{C}; 
   \label{eq:deg2-monodromy-check-3B}
\end{align}
here, $v^{[0]}_{**} = (1,2,2) \in \widetilde{\Lambda}_S^\vee$ 
and $v^{[0]}_{***} = (2,1,2)$, with the corresponding states becoming 
massless at $t^{[0]}_{**} = (-1+i)/2$ and $t^{[0]}_{***}=(-1+i)$, respectively. 
It is appropriate that the relations (\ref{eq:deg2-monodromy-check-3A}, 
\ref{eq:deg2-monodromy-check-3B}) hold without including monodromy 
contributions from the ${\rm PSL}_2\Z$ orbit of $t^{[1]}_*$ on the 
right-hand sides (see Figure~\ref{fig:cntr-gamma3-deg2}~(b)), because 
the computation (\ref{eq:PP-deg2-n12=0-result}) is for 
$f_*$ in (\ref{eq:f*-def-n1=0}), which is for $n_{1/2}=0$,  

Note that we have tools to compute the monodromy matrices $M_{\tilde{g}}$ 
from first principle, using numerical evaluation of the period polynomials. 
It is not that the relations such as (\ref{eq:deg2-monodromy-check-2}, 
\ref{eq:deg2-monodromy-check-3A}, \ref{eq:deg2-monodromy-check-3B}) 
are imposed to constrain $\Lambda_{\tilde{g}}$ that is otherwise 
intractable;\footnote{
\label{fn:imposeRltn-or-derive-Rltn}
Although Refs. \cite{AFGNT, AP} observed that the period polynomials 
$P_{\tilde{g}}$ are relevant to the $\Lambda_{\tilde{g}}$ part of the 
monodromy matrices $M_{\tilde{g}}$ (as reviewed in section \ref{sssec:PP}), 
the period polynomials $P_{\tilde{g}}$ of {\it paths} 
$\tilde{\gamma}_{\tilde{g}}$ were not exploited to compute $M_{\tilde{g}}$ there. 
When it comes to the discussion on monodromy matrices, Refs. \cite{AFGNT, AP}
used the monodromy matrices $M({\bf 1}, \Lambda_{\tilde{g}(X(v))})$ 
in (\ref{eq:monodromy-matrix-gen-shiftpart}), which follow from 4d 1-loop 
beta functions, for {\it loops} $\tilde{\gamma}_{\tilde{g}(X(v))}$ to impose 
conditions such as (\ref{eq:deg2-monodromy-check-2}, 
\ref{eq:deg2-monodromy-check-3A}, \ref{eq:deg2-monodromy-check-3B}, 
\ref{eq:U-monodromy-check-s}), and presented an example of 
$\Lambda_{\tilde{g}_i}$'s satisfying those conditions.  
} %
we computed the matrices $\Lambda_{\tilde{g}}$ from first principle and 
expressed them in terms of the (implicit $n_{1/2}=0$ and) integration 
constants $d_{111}$, $b_{1}$ and $\chi$;  
the relations (\ref{eq:deg2-monodromy-check-2}, 
\ref{eq:deg2-monodromy-check-3A}, \ref{eq:deg2-monodromy-check-3B}) 
are satisfied automatically, as explained 
in (\ref{eq:monodrm-product-compatible-PP}--\ref{eq:monodrm-product-compatible-matrix}). 

Let us now go back to the program of imposing the integrality 
of the monodromy matrices to narrow down theoretically possible 
choices of the classification invariants. Demanding 
that all the matrix entries in (\ref{eq:deg2-Lg2-result}) are integers, 
we obtain conditions that are independent from (\ref{eq:cond-from-PQ-deg2-B}). 
A common solution to those conditions is parametrized as 
\begin{align}
   \chi = -252, \quad   d_{111} = 2D, \quad
   b_1 = 8D + 36 + 24B, \quad  a_{11} = A, 
\end{align}
for $D, A, B \in \Z$. The first two conditions impose extra 
conditions on the classification invariants $(n_0, n_{1/2}, b_{\cal R})
=(-2,0,b_{\cal R})$. The required value $\chi = -252$ here 
is the same as the value $[-c_0(0)]_{n_{1/2}=0} = -252$ determined 
in an independent reasoning (\ref{eq:matching-GV+chi}), so 
the extra condition is satisfied (see also section \ref{sec:discussion}).  
The condition that $d_{111}$ is even (for $n_{1/2}=0$) implies that 
only 
\begin{align}
  b_{\cal R} \in 2\Z
   \label{eq:deg2-b-div-by-2}
\end{align}
are for branches of theoretically consistent Het--IIA dual moduli space. 

We are also ready to compare the two parameters of the 4d effective theory, 
$(b_1)_{+24\Z}$ and $(c_2)_1$ modulo $24\Z$ (not just mod $12\Z$ as in 
(\ref{eq:deg2-compare-b1-n-c21-mod12})). 
\begin{align}
  (b_1)_{+24\Z} - (c_2)_1 \equiv
    \left( 36 + 16 - 4b_{\cal R} \right) - \left( 52 -4b_{\cal R} \right)
  \equiv 0. 
  \label{eq:cond-from-PQ-gen-4Wall-open-b}
\end{align}
So, for any branch of Het--IIA dual vacua with integral monodromy matrices, 
we have seen that the two parameters $(b_1)_{+24\Z}$ and $(c_2)_1$ 
yield one common value that is interpreted as $\int_M c_2(TM) D_{a=1}$
if the branch is given by a non-linear sigma model with the target 
space $M$ in the Type IIA description. 

Now, the conditions (\ref{eq:cond-from-PQ-gen-4Wall}) is read precisely 
as Wall's condition for (necessary and) sufficient condition for 
a diffeomorphism class $[M]$ of real 6-manifolds $M$ to exist, with 
the trilinear intersection form on $H^2(M;\Z)$ and the 2nd Chern class 
in $H^4(M;\Z)$ characterized by $(C_{11}, d_{111})$ and $(24, (c_2)_1)$.  
The conditions (\ref{eq:cond-from-PQ-gen-4Wall-open-a}, 
\ref{eq:cond-from-PQ-gen-4Wall-open-b}) are equal to the additional 
conditions (\ref{eq:open-geom-cond-a}, \ref{eq:open-geom-cond-b}) 
for the central charge of 4d ${\cal N}=2$ supersymmetry 
from D-branes appropriate in a phase of non-linear sigma model description. 
So, we have seen that all the branches of theoretically consistent 
Het--IIA dual moduli space with $\Lambda_S = \vev{+2}$ and $n_{1/2} = 0$
contain phases described by a non-linear sigma model\footnote{
\label{fn:just-vect}
To be precise, what we have confirmed is existence and unique 
determination of a diffeomorphism class of real 6-dimensional 
manifolds $[M]$ that reproduces $(C_{ab}, d_{abc})$ and 
$(24, (c_2)_a)$ of a branch with $(\Lambda_S, \Lambda_T)$, 
$\{ \Phi_\gamma \}$ and $\{ \Psi_\gamma \}$, along with a few consistency 
conditions (\ref{eq:cond-from-PQ-gen-4Wall-open-a}, 
\ref{eq:cond-from-PQ-gen-4Wall-open-b}) for the branch to be realized 
by a non-linear sigma model with the target space $M$. 
We have not explored any observable from 4d hypermultiplets 
for consistency check on the geometric phase interpretation.  
} %
 in the Type IIA language. 

Now, three (or possibly more)\footnote{
\label{fn:beyond-diffeo-classify}
On the subtle possibility that something other than BPS dyon spectra
might be used for an even finer classification, see \cite{BW-16, EW19}.
} %
 distinct branches of moduli space 
remain (among those with $\Lambda_S = \vev{+2}$ and $n_{1/2} = 0$). 
The integer parameters $A$ and $B$ differ by $+\Z$ do not lead to 
distinct spectra of electrically/magnetically charged 
4d ${\cal N}=2$ BPS states (note that 
$\Delta a_{11} \in \Z$ and $\Delta b_1 \in 24\Z$ and 
remember (\ref{eq:abd-ambiguity})). $\Delta D \in 3\Z$ also 
fall into the ambiguity in (\ref{eq:abd-ambiguity}). 
So, $D_{+3\Z} \in \{2,1,0\}_{+3\Z}$ (or equivalently 
$(b_{\cal R})_{+6\Z} \in \{ 0,2,4\}_{+6\Z}$) label the three branches.
Calabi--Yau threefolds for those three branches are known in the 
literature \cite{KKRS, BW-16}.

\subsubsection{The Cases $n_{1/2} = 1,2,3,4$}

It is straightforward to employ the same method to determine the 
monodromy matrix $M_{\tilde{g}_2}$ for the cases with $\Lambda_S = \vev{+2}$ 
and $n_{1/2} > 0$. The weight-6 automorphic form $f_*$ 
in (\ref{eq:f*-def-rho=1}) under ${\rm PSL}_2\Z$ must be determined 
uniquely for individual $n_{1/2} \in \{ 0,1,2,3,4 \}$, and it should be 
of the form 
\begin{align}
 f_*(t) & \; = \frac{1}{(2\pi i)^3}
      \frac{a E_4^9 + b E_4^6 E_6^2 + c E_4^3 E_6^4 + d E_6^6}{E_4^3 E_6^3} 
  \label{eq:f*-deg2-gen-form}
\end{align}
for some coefficients $a,b,c,d$ because $\partial_t^2 {\cal F}^{(1)}$
(and hence the three-fold integral of $f_*$) should have logarithmic 
singularity at $t \simeq t^{[0]}_* = i$ (where $E_6(t) \simeq 0$) and 
also at $t \simeq t^{[1]}_* = e^{2\pi i /3}$ (where $E_4(t) \simeq 0$). 
The coefficients $a,b,c,d$ may, in principle, be determined by 
exploiting (\ref{eq:f*-KG-relatn-rho=1}, \ref{eq:KG-Phi-relatn}); 
we demand instead that the $e^{2\pi i t}$ series expansion of $f_*$ 
agrees with the $e^{2\pi i t}$ series expansion of 
$(2\pi i)^{-5}\partial_t^5 {\cal F}^{(1)}$, where the coefficients of 
the latter are determined by $n_{w,k=0} = c_{\gamma}((w,w)/2)$'s 
(see \S \ref{ssec:formula-Phi}).  It turns out that 
\begin{align}
  a = 2, \quad b = - \frac{23}{9}, \quad c = \frac{5+2n_{1/2}}{9}, \quad 
 d = - \frac{2n_{1/2}}{9}. 
   \label{eq:f*-deg2-gen-coeff}
\end{align}

The period polynomial $P_{\tilde{g}_2}(t;f_*)$ can be evaluated numerically 
for individual $n_{1/2} \in \{ 1,2,3,4\}$, just like we have done for 
the case $n_{1/2} = 0$. The results are fitted very well by the formula 
\begin{align}
 P_{\tilde{g}_2}(t;f_*) & \; \simeq
     \frac{\zeta(3)}{(2\pi i)^3} \frac{(-252+56n_{1/2})}{2}(t^4-1)
   + \frac{t^4+2t+1}{2}
   + \left( - \frac{3}{2} + \frac{n_{1/2}}{4} \right) (t^3+t). 
   \label{eq:PP-deg2-n12-gen-result}
\end{align}
This is combined with the contributions from the integration constant 
terms (\ref{eq:Qt-rho1-b}) to 
determine (\ref{eq:PP+cobndry=anal.contn.violatn}), and hence the 
matrix $\Lambda_{\tilde{g}_2}$. 
\begin{align}
  \Lambda_{\tilde{g}_2} & \; = \left( \begin{array}{cc|c}
    1 + \frac{\zeta(3)}{(2\pi i)^3}X_0 & 1 &
      \left(\frac{b_1}{24}-\frac{d_{111}}{6}-\frac{3}{2}+\frac{n_{1/2}}{4}\right) \\
   1 & 1 - \frac{\zeta(3)}{(2\pi i)^3}X_0 & 
      \left(\frac{b_1}{24}-\frac{d_{111}}{6}-\frac{3}{2}+\frac{n_{1/2}}{4}\right) \\
 \hline 
        \left(\frac{b_1}{24}-\frac{d_{111}}{6}-\frac{3}{2}+\frac{n_{1/2}}{4}\right) & 
         \left(\frac{b_1}{24}-\frac{d_{111}}{6}-\frac{3}{2}+\frac{n_{1/2}}{4}\right) & 0 
  \end{array} \right) + \Z \widetilde{C}, 
\end{align}
where 
\begin{align}
 X_0 := \chi +  252 - 56 n_{1/2}.
  \label{eq:use-chi-deg2-gen} 
\end{align}
When we use the value of $\chi$ determined by the reasoning 
reviewed in (\ref{eq:matching-GV+chi}) (see also \S \ref{ssec:formula-Phi}), 
the combination $X_0$ vanishes for all $n_{1/2}$. 

The value of $(b_1)_{+24\Z}$ is determined when we demand that all the 
entries of the matrix $\Lambda_{\tilde{g}_2}$ be integers. This 
condition on $b_1$ mod $24\Z$ and the 
conditions (\ref{eq:cond-from-PQ-deg2-B}) combined imposes 
one condition $d_{111} - n_{1/2} \in 2\Z$, or equivalently 
\begin{align}
  b_{\cal R} \in 2\Z_{\geq 0}
\end{align}
on the classification invariant, besides determining 
$(b_1)_{+24\Z}$ and $(a_{11})_{+\Z}$. One also finds that 
\begin{align}
 (b_1)_{+24\Z} = 52 -4b_{\cal R} -10 n_{1/2} + 24\Z = (c_2)_1 + 24\Z 
\end{align}
by using (\ref{eq:EW19-fit-deg2-4d}, \ref{eq:EW19-fit-deg2-4c2}). 

We have therefore seen that both $(b_1)_{+24\Z}$ and $(c_2)_1$ yield 
one common thing that can be interpreted as the 2nd Chern class of 
the target space in the Type IIA language. Now, the conditions
(\ref{eq:cond-from-PQ-gen-4Wall}) is read as the sufficient 
conditions for existence of a diffeomorphism class $[M]$ of real 6-dimensional 
manifolds whose trilinear intersection form and 
the 2nd Chern class on $H^2(M;\Z)$ agree with what we compute from 
the data $\Lambda_S$, $\{ \Phi \}$ and $b_{\cal R}$.
It is reasonable to conclude (cf footnote \ref{fn:just-vect}) 
that all those branches 
with $\Lambda_S = \vev{+2}$, $n_{1/2} \in \{0,1,2,3,4\}$ and 
$\Z$-valued monodromy matrices have a region described by the non-linear 
sigma model in the Type IIA language, with $M$ the target space. 
Table 1 of Ref. \cite{KKRS} shows that there is a known construction 
for Calabi--Yau threefolds $M$ whose diffeomorphism classes $[M]$
correspond to 
$n_{1/2}=1$, $b_{\cal R}+6\Z = \{ 0,2\}_{+6\Z}$, 
$n_{1/2}=2$, $b_{\cal R}+6\Z = \{0,2\}_{+6\Z}$, 
$n_{1/2}=3$, $b_{\cal R} + 6\Z = \{0\}_{+6\Z}$.
At this moment, it is not clear whether a geometry with 
$(n_{1/2}, b_{\cal R} +6\Z)=(1,4_{+6\Z}), (2,4_{+6\Z}), (3,2_{+6\Z}), 
(3,4_{+6\Z}), (4, \{0,2,4\}_{+6\Z})$ is simply not within the range 
of parameters of the toric ambient space scanned in \cite{KKRS}, 
or not within the category of complete intersection of a toric variety, 
or such a geometry does not exist for a reason we do not understand yet. 

\section{The Case $\Lambda_S = U$}
\label{sec:U}

As another example, let us work on the case $\Lambda_S = U$. 
Historically, the relevance of an automorphic form $f_*$ to the 
monodromy matrices has been observed in this $\Lambda_S = U$ case
for the first time \cite{deWit, AFGNT}, so the following 
presentation is inevitably 
very close to what is written in the literature. 
What we do here is to be faithful to the first principle calculation 
of the monodromy matrices to narrow down the theoretically possible 
choices of classification invariants, instead of finding relevance 
of $f_*$ in a string vacuum with a known construction. 

Let us start off by reviewing known things. On the lattice $\Lambda_S = U$, 
let us choose a basis $\{ e_\rho, e_u\}$ so that $(e_\rho, e_\rho) = 
(e_u,e_u) = 0$ and $(e_\rho, e_u) = 1$. The moduli space 
$D(\widetilde{\Lambda}_S)$ is parametrized by 
$\mho = (1, (t,t)/2, t)^T = (1, \rho u, \rho, u)^T$, where 
$t = e_\rho \rho + e_u u \in \Lambda_S \otimes \C$.  
The lattice isometry group ${\rm Isom}(\widetilde{\Lambda}_S)$ is 
(e.g., \cite{Moore-GKM, HLOY})
\begin{align}
  \left( {\rm PSL}_2^{(\rho)}\Z \times {\rm PSL}_2^{(u)}\Z \right)
    \rtimes (\Z_2\vev{\sigma} \times \Z_2(-1)_{\rho u}) \times \Z_2(-id), 
\end{align}
where ${\rm PSL}_2^{(\rho)}\Z$ and ${\rm PSL}_2^{(u)}\Z$ act on 
$D(\widetilde{\Lambda}_S) = \{ (\rho, u) \in \C \; | \; 
{\rm Im}(\rho){\rm Im}(u) > 0 \}$ as linear fractional transformations
on $\rho$ and $u$, respectively, and $(-1)_{\rho u}: (\rho, u) \mapsto 
(-\rho, -u)$. The map $\sigma$ brings $(\rho, u)$ to 
$(\rho^\sigma, u^\sigma) = (u, \rho)$. The discriminant group $G_S$ is 
trivial for $\Lambda_S = U$, and so is the group ${\rm Isom}(G_S,q_S)$. 
So, all the elements in ${\rm Isom}(\widetilde{\Lambda}_S)$ lifts to 
isometries of ${\rm Isom}({\rm II}_{4,20})$ for the embedding 
$\widetilde{\Lambda}_S = {\rm II}_{2,2} \hookrightarrow 
{\rm II}_{2,2} \oplus {\rm II}_{2,18} = {\rm II}_{4,20}$. 
For this reason, we will demand in this article that all the elements 
$g \in {\rm Isom}(\widetilde{\Lambda}_S)$ should have a lift duality 
transformation $\tilde{g}$ whose monodromy matrix is $\Z$-valued.\footnote{
It is fine to focus on $({\rm PSL}_2^{(\rho)}\Z \times {\rm PSL}_2^{(u)}\Z)
\rtimes \Z_2\vev{\tilde{\sigma}}$ for the reasons explained 
already in footnotes \ref{fn:-1-projectiveness} and \ref{fn:prd-dom-not-conn}. 
} %

Before starting to work out the monodromy matrices, let us 
also quote the results on classification invariants for the case 
$\Lambda_S = U$. The invariant $\{ \Phi \}$ is the unique
scalar valued weight $(11-\rho/2 = 10)$ modular form starting with 
$n_{\gamma =0} = -2$, which is $\Phi = -2 E_4 E_6$. One more 
classification invariant is 
\begin{align}
b_{\cal R} \in 2^{-1} \Z_{\geq 0}.
  \label{eq:para-in-Psi-U}
\end{align}
With this invariant, the parameters of the effective theory 
in the ${\rm Im}(u) > {\rm Im}(\rho)$ phase are given 
by \cite{HM, Stieberger, EW19}
\begin{align}
   d_{\rho\rho u} & \; = n'-2 + 2 \delta n_\rho, \qquad 
  d_{\rho uu}  = n' + 2 \delta n_u, \qquad
   d_{\rho\rho\rho} = 2, \qquad d_{uuu}=0, 
    \label{eq:EW19-fit-U-4d} \\
  (c_2)_\rho & \; = 12(2+n')-4 + 24 \delta n_\rho, \qquad
  (c_2)_u = 12(2+n') + 24 \delta n_u; 
   \label{eq:EW19-fit-U-4c2}
\end{align}
Here, $n' := 2-b_{\cal R}/6$. 

\subsection{Analysis on the $\Lambda_S = U$ Case}

Now, let us choose a base point of paths in a connected part 
${\cal H}_\rho \times {\cal H}_u$ of the covering space 
$D(\widetilde{\Lambda}_S)$, 
\begin{align}
t_0^{(u)} = (\rho_0^{(u)}, u_0^{(u)}) \simeq 
    (i {\rm ~large~positive~finite},  \; +i\infty);   
\end{align}
$\rho_0^{(u)}$ is chosen to be purely imaginary with the imaginary part 
finite but much larger than 1. 
All the loci of extra massless fields are \cite{AFGNT} of the form of 
$\rho = u^g$ for some $g \in {\rm PSL}_2^{(u)}\Z$, and form a single 
orbit under ${\rm Isom}(\widetilde{\Lambda}_S)$. 
Monodromy matrices for loops in the covering space is completely 
understood, so we are left to choose a set of generators $\{ g_i \}$ of 
$[{\rm PSL}_2^{(\rho)}\Z \times {\rm PSL}_2^{(u)}\Z] \rtimes \Z_2\vev{\sigma}$, 
find one lift $\tilde{g}_i$ (and $\tilde{\gamma}_{\tilde{g}_i}$) for each 
$g_i$ and compute the matrix $M_{\tilde{g}_i}$. 
As a set of generators, we choose $\{ g_{\infty(u)}^{\pm 1}, 
g_{\infty(\rho)}^{\pm 1}, \sigma, g_{2(u)}^{\pm 1}\}$, where 
$g_{\infty(u)}$ and $g_{2(u)}$ keep $\rho$ invariant and act on $u$ 
as $u \mapsto u+1$ and $u \mapsto -1/u$, respectively. Similarly, 
$g_{\infty(\rho)}: \rho \mapsto \rho + 1$. 

\begin{figure}
\begin{center}
 \begin{tabular}{ccc}
    \includegraphics[width=0.3\linewidth]{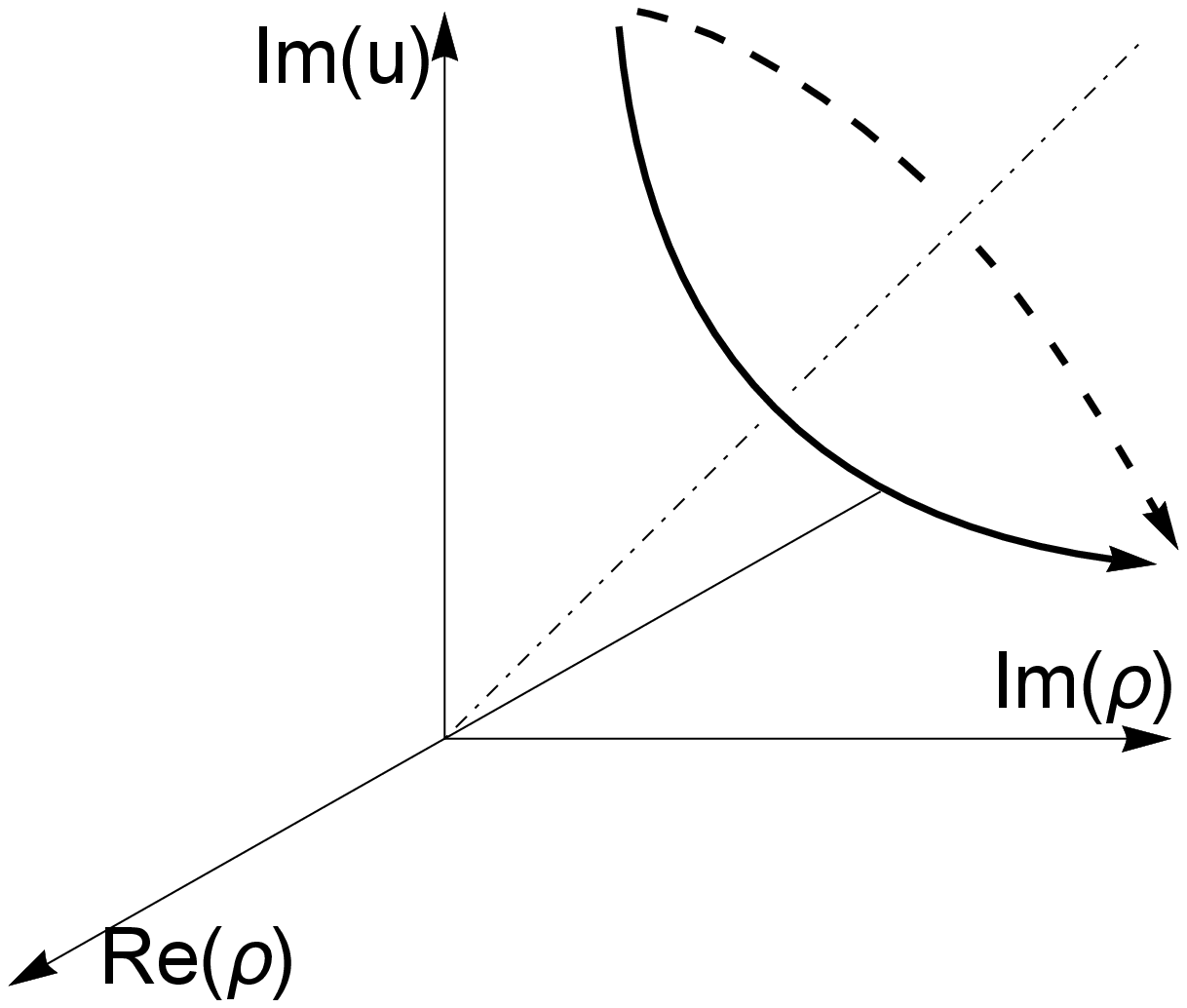} & $\quad$ &
    \includegraphics[width=0.25\linewidth]{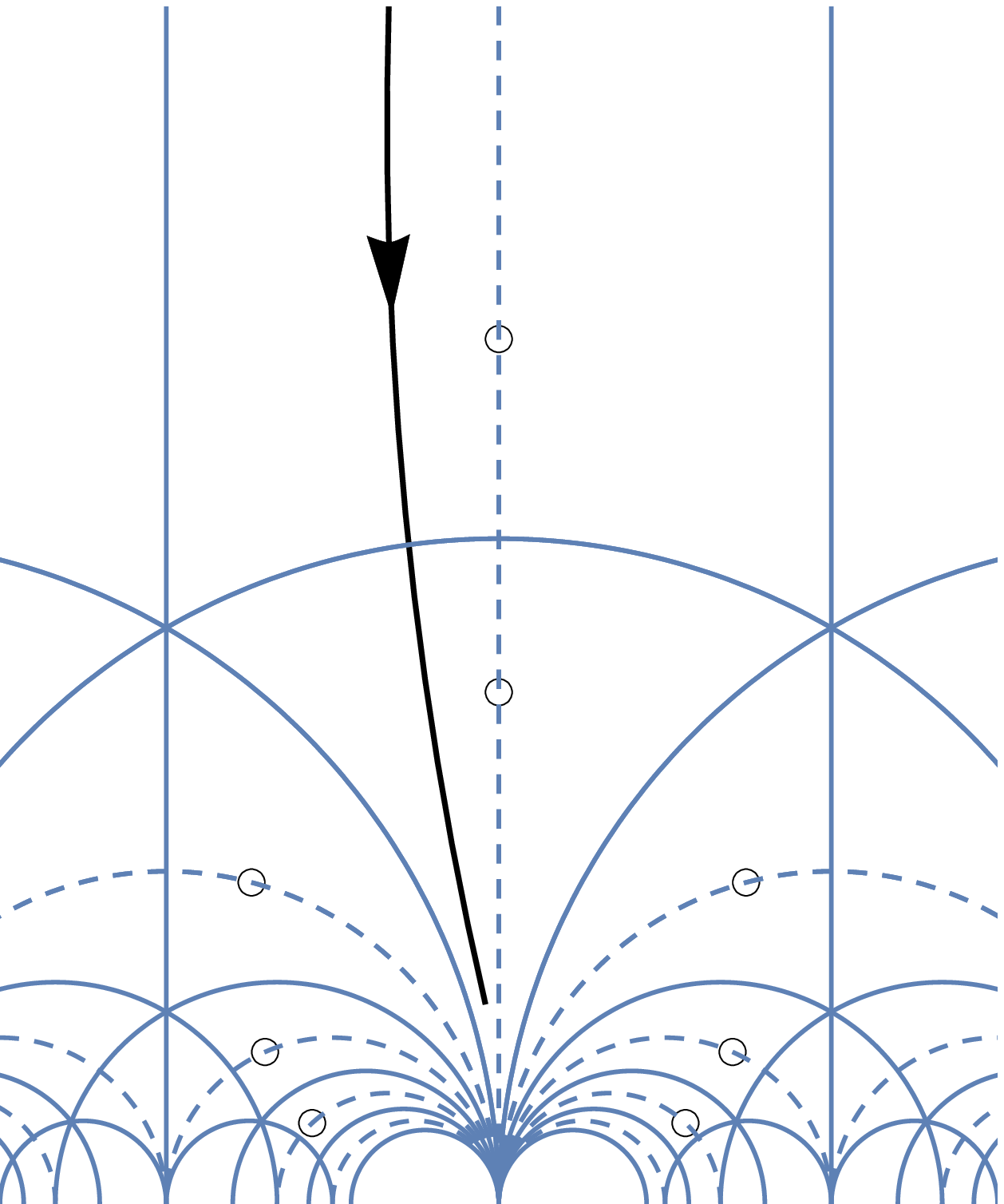} 
 \\
  (a) & & (b)
 \end{tabular}
  \caption{\label{fig:for-U} The paths $\tilde{\gamma}_{\tilde{\sigma}}$
and $\tilde{\gamma}_{\tilde{\sigma}^{-1}}$ are shown by the solid and 
dashed curves, respectively, in the ${\rm Im}(u)=0$ slice of the 
$(\rho, u) \in \C^2$ space in (a); the $u=\rho$ singularity 
is shown by the dot-dashed line. 
The panel (b) shows the path $\tilde{\gamma}_{\tilde{g}_{2(u)}}$ in the 
$\rho = \rho_0^{(u)}$ slice of the $(\rho, u)$ parameter space.  
The singularity at $u \in {\rm PSL}_2\Z \cdot \rho_0^{(u)}$ are marked by 
the open circles. }
\end{center}
\end{figure}

To specify one lift $\tilde{g}_i$ for each $g_i$, we describe 
the corresponding path $\tilde{\gamma}_{\tilde{g}_i}$, as follows. 
The paths $\tilde{\gamma}_{\tilde{g}_{\infty(\rho)}}$ and 
$\tilde{\gamma}_{\tilde{g}_{\infty(u)}}$ both start from $t_0^{(u)}$, 
and have $\Delta \rho = +1$ for fixed $u=u_0^{(u)}$ and 
$\Delta u = +1$ for fixed $\rho = \rho_0^{(u)}$, respectively, to reach 
the endpoint $(t_0^{(u)})^{g_{\infty(\rho)}}$ and $(t_0^{(u)})^{g_{\infty(u)}}$, 
respectively. 
The path $\tilde{\gamma}_{2(u)}$ starts from $t_0^{(u)}$, and moves in 
the $u$-plane down the imaginary axis (and fixed $\rho = \rho_0^{(u)}$) 
to $u = -1/(u_0^{(u)})$, while avoiding $u = \rho_0^{(u)}$ by detouring 
into the 2nd quadrant in the $u$-plane. Finally, the path 
$\tilde{\gamma}_{\tilde{\sigma}}$ [resp. $\tilde{\gamma}_{\tilde{\sigma}^{-1}}$]
starts from $t_0^{(u)}$ and ends at $(t_0^{(u)})^\sigma = (u_0^{(u)}, \rho_0^{(u)})$, avoiding the $u=\rho$ singularity on the way by temporarily setting 
${\rm Re}(\rho)$ 
to be positive [resp. negative]; see Figure \ref{fig:for-U}~(a).
The corresponding duality transformation (analytic continuation of the 
section $\Pi = (X, F)^T$ along those paths) are denoted by 
$\tilde{g}_{\infty, \rho}$, $\tilde{g}_{\infty, u}$, $\tilde{g}_{2,u}$, and 
$\tilde{\sigma}^{\pm 1}$, respectively. 

The monodromy matrices for $\tilde{g}_{\infty(\rho)}$ and 
$\tilde{g}_{\infty(u)}$ are given by $g_{\infty(\rho)}$, $g_{\infty(u)}$
and (\ref{eq:Lambda-infty-formula}). The effective theory parameters 
are subject to the conditions (\ref{eq:cond-from-PQ-gen-4Wall}, 
\ref{eq:cond-from-PQ-gen-4Wall-open-a}), or more explicitly,  
\begin{align}
&  d_{\rho\rho\rho}, \quad d_{\rho \rho u}, \quad d_{\rho uu}, \quad d_{uuu} \in \Z, 
     \label{eq:cond-ginf-Z-4U-1} \\
&  (a_{\rho\rho})_{+\Z} = (d_{\rho\rho\rho}/2) + \Z, \quad
  (a_{uu})_{+\Z} = (d_{uuu}/2) + \Z, \label{eq:cond-ginf-Z-4U-2} \\
&  (a_{\rho u})_{+\Z} = d_{\rho\rho u}/2 +\Z, \label{eq:cond-ginf-Z-4U-3} \\
&  d_{\rho\rho u} \in  d_{\rho uu} + 2\Z, \label{eq:cond-ginf-Z-4U-4} \\
&     2d_{uuu} + b_u \in 12\Z,  \qquad 2d_{\rho\rho\rho} + b_\rho \in 12\Z.
  \label{eq:cond-ginf-Z-4U-5}
\end{align}
Those conditions imply that the classification invariant $b_{\cal R}$
can take values only in 
\begin{align}
  b_{\cal R} \in 6 \Z_{\geq 0}
\end{align}
in a branch of theoretically consistent Het--IIA dual moduli space, 
not $2^{-1} \Z_{\geq 0}$. The values of $(a_{\rho\rho})_{+\Z}$ and 
$(a_{uu})_{+\Z}$ are also fixed, when combined with the 
parametrization (\ref{eq:EW19-fit-U-4d}). The values of 
$(b_\rho)_{+24\Z}$ and $(b_u)_{+24\Z}$ are determined only 
modulo $+12\Z$. At this level of precision, 
\begin{align}
  (b_\rho)_{+12\Z} = (-2d_{\rho\rho\rho})_{+12\Z} = -4 + 12\Z, 
    &  \qquad 
  (c_2)_\rho \equiv -4 \quad ({\rm mod~}12\Z), \\
  (b_u)_{+12\Z} = (-2d_{uuu})_{+12\Z} = 12\Z, & \qquad 
  (c_2)_u \equiv 0 \quad ({\rm mod~}12\Z). 
\end{align}
So, we are left with $(b_a)_{+24\Z} \equiv (c_2)_a$ mod 24, or 
$(c_2)_a+12$ mod 24, chosen individually for $a=\rho$ and $a=u$.  

The monodromy representation matrices $M_{\tilde{\sigma}^{\pm 1}} = 
M(\sigma, \Lambda_{\tilde{\sigma}^{\pm 1}})$ are given by computing 
the analytic continuation of ${\cal F}^{(1)}$ from 
$t_0 = (\rho^{(u)}_0, u^{(u)}_0)$ to $t_0^{\sigma} = (u_0^{(u)}, \rho_0^{(u)})$ 
along the paths $\tilde{\gamma}_{\tilde{\sigma}^{\pm 1}}$. 
In addition to the polynomial terms of ${\cal F}^{(1)}$, one more term 
$c(-1){\rm Li}_3(e^{2\pi i (u-\rho)})/(2\pi i)^3$ in ${\cal F}^{(1)}$ contributes 
to ${\cal F}^{(1)}(t^{\tilde{\sigma}^{\pm 1}})-{\cal F}^{(1)}(t)$ in 
(\ref{eq:deWit-1loopSuper-constraint}). Using\footnote{
The numerator in the 1st term is the function of $(\rho, u)$ 
whose value is ${\rm Li}_{3}(e^{2\pi i (u'-\rho')})_{(\rho', u') 
= (\rho^\sigma, u^\sigma)}$; the value is determined by analytically 
continuing ${\rm Li}_{3}(e^{2\pi i (u'-\rho')})$ for $(\rho', u')$
along $(\tilde{\gamma})^\sigma \circ \tilde{\gamma}_{\tilde{\sigma}^{\pm 1}} \circ 
(\tilde{\gamma})^{-1}$. 
} %
\begin{align}
  \frac{\left[{\rm Li}_3(e^{2\pi i (u-\rho)})\right]_{
      {\rm along~}\tilde{\gamma}_{\tilde{\sigma}^{\pm 1}}}}{(2\pi i)^3} - 
  \frac{{\rm Li}_3(e^{2\pi i (u-\rho)})}{(2\pi i)^3} = \frac{1}{3!}\left[
   (u-\rho)^3 \mp \frac{3}{2}(u-\rho)^2 +\frac{1}{2} (u-\rho)
       \right] 
\end{align}
and $c(-1) = -2$ (see \S \ref{ssec:formula-Phi}), one finds that 
\begin{align}
 {\cal F}^{(1)}(t^{\tilde{\sigma}^{\pm 1}}) - {\cal F}^{(1)}(t) & \; = 
    \frac{\pm 1 -a_{\rho\rho}+a_{uu}}{2}u^2
  + \frac{\pm 1 + a_{\rho\rho}-a_{uu}}{2} \rho^2
   \mp \rho u + \frac{4+b_\rho-b_u}{24}(\rho -u); 
 \label{eq:continatn-by-sigma}
\end{align}
cubic terms cancel when we use (\ref{eq:EW19-fit-U-4d}). 
The monodromy matrices $M_{\tilde{\sigma}} = M(\sigma,\Lambda_{\tilde{\sigma}})$
and $M_{\tilde{\sigma}^{-1}} = M(\sigma, \Lambda_{\tilde{\sigma}^{-1}})$ 
are  now determined by (\ref{eq:deWit-1loopSuper-constraint}). 
\begin{align}
  \sigma = \left[ \begin{array}{cc|cc} 
     1 & & & \\ & 1 & & \\ \hline
     & & & 1 \\ & & 1 & \end{array} \right],  \quad 
  \Lambda_{\tilde{\sigma}^{\pm 1}} = \left[ \begin{array}{cc|cc}
    0 & 0 & \frac{4+b_\rho-b_u}{24} & -\frac{4+b_\rho-b_u}{24} \\
    0 & 0 & 0 & 0 \\
  \hline 
      & & \pm 1+a_{\rho\rho}-a_{uu} & \mp 1 \\
      & & \mp 1 & \pm 1-a_{\rho\rho}+a_{uu} \end{array} \right]
    + \Z \widetilde{C} ; 
  \label{eq:result-U-Ls}
\end{align}
the lower-left corner of the symmetric matrices $\Lambda_{\tilde{\sigma}^{\pm 1}}$ 
are omitted just to save space. The two matrices automatically satisfy 
the relations (cf footnotes \ref{fn:factor2-vs-AFGNT}, 
\ref{fn:imposeRltn-or-derive-Rltn} and \ref{fn:compare-MMtrcs-wAFGNT})
\begin{align}
(M_{\tilde{\sigma}^{\pm 1}})^2 \in M({\bf 1},
  \pm \Lambda_{\tilde{g}(X(v_{\rho u}))}) \cdot (M_D)^{n}, \qquad n \in \Z, 
  \label{eq:U-monodromy-check-s} 
\end{align}
where $ \pm v_{\rho u}= \pm (0,0,-1,1) \in \widetilde{\Lambda}_S^\vee$ are 
the charges of the states that become massless at $u=\rho$. 

By demanding that the matrices $\Lambda_{\tilde{\sigma}^{\pm 1}}$ are 
$\Z$-valued, we obtain one more condition. 
\begin{align}
  (b_\rho)_{+24\Z} + 4 = (b_u)_{+24\Z};  
  \label{eq:brho=bu-sigma-monodrm}
\end{align}
two choices are left, $b_\rho + 4 \equiv b_u \equiv 0$ mod 24 and 
$\equiv 12$ mod 24. 

Finally, let us determine the monodromy matrix of the duality 
transformation $\tilde{g}_{2(u)}$.  To find out how 
${\cal F}^{(1)}(\rho_0^{(u)}, -1/u_0^{(u)})$ is related to 
${\cal F}^{(1)}(\rho_0^{(u)}, u_0^{(u)})$, we use numerical 
evaluation of the period polynomial. 

For $g_{(u)} \in {\rm PSL}_2^{(u)}\Z$ mapping $u \mapsto (au+b)/(cu+d)$, 
the relation (\ref{eq:deWit-1loopSuper-constraint}) reads 
\begin{align}
 (cu+d)^2 {\cal F}^{(1)}(\rho, u^{\tilde{g}_{(u)}}) = {\cal F}^{(1)}(\rho, u)
   + \frac{1}{2} \mho(\rho, u)^T \Lambda_{\tilde{g}_{(u)}} \mho(\rho, u). 
  \label{eq:deWit-1loopSuper-constraint-U}
\end{align}
The 3rd derivative of ${\cal F}^{(1)}$ with respect to $u$ 
transforms as a modular form under ${\rm PSL}_2^{(u)}\Z$ of weight 
$(w=+4)$ \cite{deWit}, so we set 
\begin{align}
   f_*(u;\rho) := \frac{1}{(2\pi i)^3} \partial_u^3 {\cal F}^{(1)}(\rho, u). 
\label{eq:f*-def-U}
\end{align}
This modular form $f_*(u;\rho)$ is determined uniquely by the unique 
$\Phi$ (independent of the classification invariant $b_{\cal R} \in 6\Z$), 
because \cite[(2.15)]{AFGNT} (note also (\ref{eq:KG-Phi-relatn})) 
\begin{align}
 - \rho_2^2 [u_2^{-2} \partial_u u_2^2]
   \partial_u
  \bar{\partial}_{\bar{\rho}} \left( \hat{K}^{(0)a\bar{b}} G^{(1)}_{a\bar{b}} \right)
  = \cdots = \frac{i}{2} \partial_u^3 {\cal F}^{(1)}(\rho,u) = \frac{i}{2}(2\pi i)^3 f_*(u; \rho).
 \label{eq:f*-KG-relatn-U}
\end{align}
It is known \cite{deWit, AFGNT} 
that this uniquely determined $f_*(u;\rho)$ for all the $\Lambda_S=U$ 
cases is of the form (cf \cite{HKTY})
\begin{align}
  f_*(u;\rho) & \; = \frac{1}{(2\pi i)^3}\frac{-i}{\pi}
 \frac{j_u(u)}{j(u)-j(\rho)} \frac{j(\rho)}{j(u)}
   \frac{j_u(u)}{j_\rho(\rho)}  \frac{j(\rho)-j(i)}{j(u)-j(i)}, \\
  & \; =  \frac{1}{(2\pi i)^3} \frac{2}{j(\rho)-j(u)} E_4(u)
     \frac{E_4E_6}{\eta^{24}}(\rho). \label{eq:f*-forU}
\end{align}
Note that $f_*(u, \rho)$ for a fixed value of $\rho$ has the behavior 
$f_*(u) \simeq e^{2\pi i u}\times ({\rm const}.) + {\cal O}(e^{2\pi i 2 u})$ 
at large ${\rm Im}(u)$, so this is a cusp form of weight $(w=4)$ with 
poles (at $u = \rho$ and its ${\rm PSL}_2\Z$ images) in the upper half 
plane of $u$. 

The part of the prepotential ${\cal F}^{(1)}(\rho, u)$ should be 
reproduced from $f_*(u;\rho)$ by a $(w-1)=3$-fold integral with respect 
to the coordinate $2\pi i u$. The indefinite integral (the Eichler integral) 
$I[u;f_*,\tilde{\gamma},\rho] 
= F_{\rm Eich}(u;f_*,\tilde{\gamma},\rho)$ starting from 
$u=u_0^{(u)}\simeq i\infty$ converges (see the review in 
section \ref{sssec:PP}), and makes sense uniquely when 
the path $\tilde{\gamma}$ stops at ${\rm Im}(u) \gg {\rm Im}(\rho)$; 
for this reason, the $(w-1)$-fold integral (\ref{eq:add-intg-cnst-2EichlerI}) 
is ready for an easy interpretation in the ${\rm Im}(u) > {\rm Im}(\rho)$ 
phase. The integration constant terms $Q(u;\rho)$ in 
(\ref{eq:add-intg-cnst-2EichlerI}) is a polynomial of $u$ of at most 
degree $(w-2) = 2$, which should be of the form 
\begin{align}
 Q(u;\rho) & \; = \left( \frac{d_{\rho uu}}{2}\rho - \frac{a_{uu}}{2}\right) u^2
 + \left( \frac{d_{\rho\rho u}}{2}\rho^2 - a_{\rho u} \rho - \frac{b_u}{24}\right) u
   \\
& \quad 
  + \left( \frac{d_{\rho\rho\rho}}{6}\rho^3 - \frac{a_{\rho\rho}}{2}\rho^2 - \frac{b_\rho}{24} - \frac{\zeta(3)}{(2\pi i)^3}\frac{\chi}{2}
   + \sum_{n\geq 1}\frac{n_{(n,0),0}}{(2\pi i)^3}{\rm Li}_3(e^{2\pi i \rho n}) \right).
   \nonumber 
\end{align}
Neither the Eichler integral $F_{\rm Eich}(u;f_*,\tilde{\gamma},\rho)$ 
nor the integration constant terms give rise to the $u^3$ term 
in the ${\rm Im}(u) > {\rm Im}(\rho)$ phase, but that is consistent 
with the known 
parametrization $d_{uuu}=0$ (for ${\rm Im}(u) > {\rm Im}(\rho)$) by 
the classification invariants (see (\ref{eq:EW19-fit-U-4d})).  

Now let us determine $\Lambda_{\tilde{g}_{2(u)}}$ in the 
monodromy matrix $M_{\tilde{g}_{2(u)}} = M(g_{2(u)},\Lambda_{\tilde{g}_{2(u)}})$, 
by using (\ref{eq:deWit-1loopSuper-constraint-U}, 
\ref{eq:PP+cobndry=anal.contn.violatn}). We evaluated the period polynomial
$P_{\tilde{g}_{2(u)}}(u;f_*,\rho)$ for several values of $\rho$ by carrying out 
numerical integrals just like in the $\Lambda_S=\vev{+2}$ cases. 
It turns out that there is a nice fitting formula for the numerical 
integrals. The $u$ and $(u^2+1)$ terms in the polynomial 
$P_{\tilde{g}_{2(u)}}(u;f_*,\rho)$ are 
\begin{align}
 [P_{\tilde{g}_{2(u)}}(u;f_*,\rho)]_{u,(u^2+1)} & \; \simeq -2(\rho^2+1)u
  + \frac{\rho^2+1}{2}(u^2+1), 
  \label{eq:PP-U-result-u-u2+1}
\end{align}
and the $(u^2-1)$ term satisfies 
\begin{align}
 [P_{\tilde{g}_{2(u)}}(u;f_*,\rho)]_{(u^2-1)}
 & \; + (u^2-1) \left[ \frac{\zeta(3)}{(2\pi i)^3} \frac{480}{2}
   + \frac{480}{(2\pi i)^3} \sum_{n\geq 1} {\rm Li}_3(e^{2\pi i \rho n}) \right]
  \nonumber  \\
 & \; \simeq (u^2-1)\left[ \frac{5}{6} \rho - \frac{1}{3}\rho^3\right], 
   \label{eq:PP-U-result-u2-1}
\end{align}
where we have used $\chi = - c(0) = -480$ and 
$n_{(n,0),0}=c(n\cdot 0) = c(0) = 480$ from (\ref{eq:matching-GV+chi})
(see \S \ref{ssec:formula-Phi}). 
So, the deviation (\ref{eq:PP+cobndry=anal.contn.violatn}) from the 
modular transformation property for ${\cal F}^{(1)}(\rho, u)$ 
is now purely polynomial in both $\rho$ and $u$. Moreover, 
the $(u^2-1)\rho^3$ terms cancel because $d_{\rho\rho\rho}=2$ in 
(\ref{eq:EW19-fit-U-4d}). So, we obtain 
\begin{align}
\Lambda_{\tilde{g}_{2(u)}} = \left[ \begin{array}{cc|cc}
   1-a_{uu} & 0 & \frac{b_\rho +12 d_{\rho uu}-20}{24} & -2+\frac{b_u}{12} \\ 
           & 1-a_{\rho\rho} & -2-d_{\rho\rho u} & 
                                  - \frac{b_\rho +12 d_{\rho uu}-20}{24}  \\
  \hline 
    & & 1+a_{\rho\rho} & 2a_{\rho u} \\
    & & & 1+a_{uu} 
  \end{array} \right] + \Z \widetilde{C}. 
\end{align}

Almost all the entries of the matrix $\Lambda_{\tilde{g}_{2(u)}}$ are 
automatically integers based on the conditions that have been derived. 
One new condition is obtained, however, which determines $(b_\rho)_{+24\Z}$
uniquely, and consequently also determines $(b_u)_{+24\Z}$ because 
of (\ref{eq:brho=bu-sigma-monodrm}). 
\begin{align}
  (b_\rho)_{+24\Z} = 20 + 12 n'+24\Z, & \qquad (c_2)_\rho \equiv 20 + 12n'
       \quad ({\rm mod~}24\Z), \\
  (b_u)_{+24\Z} = (b_\rho)_{+24\Z}, & \qquad (c_2)_u \equiv (c_2)_\rho + 4 
       \quad ({\rm mod~}24\Z). 
\end{align}
So the uniquely determined $(b_a)_{+24\Z}$'s agree with $(c_2)_a$ mod $24\Z$. 

Now that we have confirmed that $(b_a)_{+24\Z}$ and $(c_2)_a$'s allow 
a common thing interpreted as the 2nd Chern class, we can regard 
the conditions (\ref{eq:cond-ginf-Z-4U-1}, \ref{eq:cond-ginf-Z-4U-4}, 
\ref{eq:cond-ginf-Z-4U-5}) as Wall's necessary and sufficient condition 
for existence of a diffeomorphism class $[M]$ of real 6-dimensional 
manifolds, 
with the trilinear intersection of $H^2(M;\Z)$ and the 2nd Chern class 
given by $d_{abc}$ and $(c_2)_a$, respectively. In any one of the branches 
of Het--IIA moduli space with $\Lambda_S = U$, therefore, there is 
a region described by a non-linear sigma model with the target space $M$
in the Type IIA language (cf footnote \ref{fn:just-vect}). 
The properties (\ref{eq:cond-ginf-Z-4U-2}, 
\ref{eq:cond-ginf-Z-4U-3}) and $(b_a) \equiv (c_2)_a$ mod 24 guarantee 
that necessary conditions (\ref{eq:open-geom-cond-a}, 
\ref{eq:open-geom-cond-b}) for D-brane central charges in a geometric 
phase are satisfied. 

Those branches of moduli space with $\Lambda_S=U$ come with the unique 
$\Phi$ and the invariant $b_{\cal R} \in 6\Z_{\geq 0}$. Two $b_{\cal R}$'s 
that differ by $12\Z$ (i.e., two $n'$'s that differ by $2\Z$) result in 
an identical spectrum of 4d ${\cal N}=2$ BPS dyons. The two branches 
with distinct dyon spectra,\footnote{
\label{fn:compare-MMtrcs-wAFGNT}
A set of monodromy matrices of generators of 
$({\rm PSL}_2^{(\rho)}\Z \times {\rm PSL}_2^{(u)}\Z)\rtimes \Z_2(\vev{\sigma}$
is presented in \cite[(4.16)]{AFGNT}. The matrix 
$\tilde{c}/2$ for $\sigma$ in \cite[(4.16)]{AFGNT}
is the closest to $\Lambda_{\tilde{\sigma}^{+1}} + \widetilde{C}/2$ 
in this article, but they are not equal no matter how we set the value 
of $a_{\rho\rho}$, $a_{uu}$, $b_\rho$ and $b_u$ in (\ref{eq:result-U-Ls}). 
Either the choices of the symplectic 
frame at the base point $t_0$ are different (not just for 
$\Delta a_{ab} \in \Z$ and $\Delta b_a \in 24\Z$) 
between \cite{AFGNT} and here, or the monodromy matrix for $\sigma$ 
in \cite[(4.16)]{AFGNT} is not precisely for the path 
$\tilde{\gamma}_{\tilde{\sigma}}$ but for some other path in the set 
of $(loops) \circ \tilde{\gamma}_{\tilde{\sigma}} \circ (loops)$, 
where ``loops'' refer to the loops in the covering space 
$({\cal H}_s/\Z D) \times (D(\widetilde{\Lambda}_S) \backslash X_{\rm singl})$.  
We did not try to find out whether the matrices presented 
in \cite[(4.16)]{AFGNT} are for the $b_{\cal R}\equiv 0$ (12) branches, 
or for the $b_{\cal R} \equiv 6$ (12) branch. 
} %
 one with $b_{\cal R} \equiv 0$ (12) 
and the other with $b_{\cal R} \equiv 6$ (12) (i.e., even $n'$ and odd $n'$)
have known descriptions both in the Type IIA and Heterotic languages. 
In Type IIA, the corresponding Calabi--Yau threefold is the elliptic
fibration over the Hirzebruch surface $F_0$ (or $F_2$) and $F_1$, 
respectively. In Heterotic language, the internal space is K3 $\times T^2$
and the 24 instantons on K3 are distributed by $12+12$ (or $10+14$) and 
$11+13$, respectively (cf footnote \ref{fn:beyond-diffeo-classify}). 

\section{Discussion}
\label{sec:discussion}

In this article, we have given a proof of concept 
of computing the monodromy matrices directly to narrow down 
possible choices of classification invariants of Heterotic--IIA 
dual vacua; sometimes the computation involves numerical 
evaluation of period polynomials. Besides the obvious directions 
going beyond a proof of concept, there are a few questions 
of theoretical (mathematical) nature about period polynomials, 
which we note here. 

One is about the $\zeta(3) \chi/[2(2\pi i)^3]$ term in the 
prepotential. The parameter $\chi$ in the effective theory 
prepotential (\ref{eq:prepot-Het-weak}) is determined,  
through (\ref{eq:matching-GV+chi}) on one hand (which is 
based on computation of $G^{(1)}$ through $\tilde{I}[F]$ 
in (\ref{eq:def-HM-def-I})), and also 
through the imaginary coefficients in the period polynomial 
(\ref{eq:PP-deg2-n12=0-result}, \ref{eq:PP-deg2-n12-gen-result}) 
and (\ref{eq:PP-U-result-u2-1}) on the other.  
In both reasonings, the parameter $\chi$ is determined from 
the data $\{ \Phi_\gamma \}$. If a given $\{ \Phi_\gamma \}$ is 
for a theoretically consistent branch of the Heterotic--IIA dual moduli 
space, then the two procedures should result in the same value 
of $\chi$. At this moment, however, the authors do not have an idea 
how to prove mathematically that the two procedures yield the 
same value of $\chi$ for any $\{ \Phi_\gamma \}$. 

The other is about the overall transcendental factors of period 
polynomials. Even for a Hecke-eigen cusp form $f$ with integral Fourier 
coefficients that is without a pole in the interior of the complex 
upper half plane, the period 
polynomial $P_{g_2}(\sigma, f)$ do not always have coefficients 
in $\Q$; the polynomial $P_{g_2}(\sigma, f)$ may be split into 
even-power terms and odd-power terms, and then the ratio 
of coefficients among the even-power terms and also the 
ratio of those among the odd-power terms are known to be 
in $\Q$, but there is a common factor for the even-degree 
terms and another common factor for the odd-degree terms, 
which are not even necessarily algebraic (they are given by 
special values of $L$-functions, more general than the special 
values of the zeta function) \cite{Lang, Zagier-comm-trans-factor}. In the 
applications in this article, $f=f_*$ is chosen from a more general class, 
in that $f_*$ has poles in the upper half plane, and is not guaranteed 
to be a Hecke eigenform. In light of these general expectations 
for period polynomials, the fact that the coefficients of the odd degree
terms in (\ref{eq:PP-deg2-n12-gen-result}, \ref{eq:PP-U-result-u-u2+1}, 
\ref{eq:PP-U-result-u2-1}) are in $\Q$---not just their ratios 
are---hints that there are still things to be understood. 
For $\{ \Phi_\gamma \}$ to be for a theoretically consistent 
branch of the Heterotic--IIA dual moduli space, the overall 
transcendental factors of the odd part of the period polynomial 
$P_{\tilde{g}}(t,f_*)$ cannot be transcendental, or even algebraic 
outside of $\Q$.  
Either there are more math to be understood, or only finite number 
of $\{ \Phi_\gamma \}$'s have the overall transcendental factor
in $\Q$ in the odd part and are for the theoretically consistent 
dual moduli space, we do not speculate in this article. 

 \subsection*{Acknowledgments}

We thank I. Antoniadis for kindly explaining derivations in 
Ref. \cite{AGNT-threshold} to one of the authors. 
We also thank Y. Sato for discussions. 
The study in this article was supported in part by 
JSPS Fellowship for Young Scientists (YE), 
Leading Graduate School FMSP program (YE), 
the brain circulation program (TW),  
a Grant-in-Aid for Scientific Research on Innovative Areas no. 6003 (TW),
and by the WPI program (YE, TW), all from MEXT, Japan. 


  \appendix 

\section{Brief Notes on Modular Forms}
\label{sec:mod-form}

In this appendix, we collect some conventions and facts 
from the literatures for convenience of readers. 

\subsection{Eisenstein Series etc.}

\paragraph{Eisenstein series}
\begin{align}
E_2 =& 1 - 24 \left( q + 3 q^2 + \cdots \right)
    = 1 - 24 \sum_{n=1}^\infty q^n \sigma_1(n)
    = 1 - 24 \sum_{m=1}^\infty \frac{m q^m}{1-q^m}, \\
E_4 =& 1 + 240\left( q + 9 q^2 + \cdots \right)
   = 1 + 240 \sum_{n=1}^\infty q^n \sigma_3(n),
   = 1 + 240 \sum_{m=1}^\infty \frac{m^3 q^m}{1-q^m}, \\
E_6 =& 1 - 504\left( q + 33 q^2 + \cdots \right)
   = 1 - 504 \sum_{n=1} q^n \sigma_5(n)
   = 1 - 504 \sum_{m=1} \frac{m^5 q^m}{1-q^m},
\end{align}
where $q = e^{2 \pi i \sigma}$, with the argument $\sigma$ taking value 
in the complex upper half plane ${\cal H}$. The Eisenstein series 
$E_4(\sigma)$ and $E_6(\sigma)$ are modular forms of weight 4 and 6, 
respectively,
for ${\rm SL}(2;\Z)$, but $E_2(\tau)$ is not modular (it is Mock 
modular).\footnote{
We will also use $\hat{E}_2(\sigma) := E_2(\sigma) - 3/(\pi {\rm Im}(\sigma))$
in (\ref{eq:def-HM-def-I}).
} %
The space of scalar-valued modular forms can be identified with the polynomial
ring $\C[E_4,E_6]$. 

In the main text, we will also use the Dedekind Eta function $\eta$ and 
the $j$-invariant
\begin{align}
    \eta := q^{1/24} \prod_{n = 1}^\infty (1-q^n), \qquad 
   j := \frac{E_4^3}{\eta^{24}}, 
\end{align}
which are of weight $+1/2$ and $0$, respectively. There is 
a relation $E_4^3- E_6^2 = 1728 \eta^{24}$. 

\paragraph{Ramanujan--Serre derivative}

For a modular form $F$ of weight $k$, the \textit{Ramanujan--Serre derivative}
$\partial^S$ is defined by
\begin{align}
 \partial^S F :=
  \left(\frac{1}{2\pi i} \frac{\partial}{\partial \sigma} - \frac{k}{12} E_2
  \right) F = \eta^{2k} q \partial_q \left(\frac{F}{\eta^{2k}} \right).
\end{align}
$\partial^S F$ is a modular form of weight $k+2$.
It is useful to know that $\partial^S E_4 = (-1/3)E_6$ 
and $\partial^S E_6 = (-1/2) E_4^2$. 

\subsection{Explicit Formulae of the Modular Forms $\{ \Phi_\gamma \}$}
\label{ssec:formula-Phi}

We will use some of the Fourier coefficients of 
$\{ \Phi_\gamma \}_{\gamma \in G_S}$ in the main text, so here is the data. 
For more information, readers are referred to the literatures cited 
in \cite{EW19}. 

\paragraph{The case $\Lambda_S = U$:}

In this case, the group $G_S$ is trivial, and $\{\Phi_\gamma \}$ consists 
of single component $\Phi$, which is equal to $n_0 E_4 E_6 = -2 E_4E_6$, 
without any free parameters.
\begin{align}
  F = \frac{\Phi}{\eta^{24}} 
   = \frac{-2}{q} + 480 + 282888 q+ 17058560 q^2+ \cdots; 
\end{align}
the Fourier coefficients $c(-1)=-2$ and $c(0)=480$ are used 
in (\ref{eq:continatn-by-sigma}, \ref{eq:PP-U-result-u2-1})
along with (\ref{eq:matching-GV+chi});
$c(1) = 282888$ and $c(2)=17,058,560$ can also be used to 
verify that $f_*$ in (\ref{eq:f*-forU}) satisfies 
(\ref{eq:f*-def-U}, \ref{eq:matching-GV+chi}). 

\paragraph{The case $\Lambda_S = \vev{+2}$:}

In this case, $G_S \cong \Z_2$, so $\{ \Phi_\gamma \}$ consists of 
two components, $\Phi_0$ and $\Phi_{1/2}$. It is known that \cite{KKRS, MP, HK}
\begin{align}
 \Phi_0 & \; = -2 + (300-56n_{1/2})q + (217200-13680 n_{1/2})q^2
               + \cdots  \nonumber \\
 \Phi_{1/2} & \; =  n_{1/2} q^{\frac{1}{4}} + (2496 + 360 n_{1/2}) q^{\frac{5}{4}}
      + (665600 + 30969 n_{1/2}) q^{\frac{9}{4}} + \cdots;
\end{align}
we have chosen a parametrization so that the coefficients of the 
leading terms of $\Phi_0$ and $\Phi_{1/2}$ are $n_0 = -2$ and $n_{1/2}$, 
respectively. 
It is straightforward to compute $F_\gamma = \Phi_\gamma/\eta^{24}$. 
Let us just note that 
\begin{align}
& c_0(0) = 252 -56 n_{1/2}, \quad 
 c_{1/2}(1/4) = 2496+384 n_{1/2}, \quad 
 c_0(1) = 223,752 -15,042 n_{1/2}, \\
& c_{1/2}(9/4) = 38,637,504 + 1,129,856 n_{1/2}, \quad
  c_0(4) = 9,100,224,984 - 115,446,576 n_{1/2}.  \nonumber  
\end{align}
Those facts are used in (\ref{eq:use-chi-deg2-gen}, \ref{eq:matching-GV+chi}) 
and also in determining $f_*$ 
in (\ref{eq:f*-def-n1=0}, \ref{eq:f*-deg2-gen-form}, 
\ref{eq:f*-deg2-gen-coeff}). 

\section{Eichler Cohomology and Coarse/Fine Classification}
\label{sec:Eichler-coh}

Here we leave a brief note on how the Eichler cohomology is 
related to the coarse versus fine classifications of branches 
of the Heterotic--IIA dual moduli space discussed in \cite{EW19}. 
The following discussion is written for the cases with 
${\rm rank}(\Lambda_S)=1$; some kinds of generalization for 
higher-rank cases may be possible, but we will not try to 
cover the higher rank cases here. 

The Eichler cohomology is often formulated in the literatures 
as follows. For a group $G$ such as ${\rm PSL}_2\Z$, 
$\Gamma_0(k)_+$ and their finite index subgroups that acts on 
the complex upper half plane ${\cal H}$, and for a cusp form $f$ of 
weight $w$ under the group $G$, the period polynomials can be 
used to think of an assignment 
\begin{align}
  P_f: G \ni g \longmapsto P_g(t, f) \in \C[t]_{{\rm deg}=w-2}. 
  \label{eq:assign-PP-holCase}
\end{align}
On the abelian group $\C[t]_{{\rm deg}=w-2}$ of the $\C$-coefficient 
polynomials of degree $(w-2)$, the group $G$ acts as 
\begin{align}
 \C[t]_{{\rm deg}=w-2} \ni \chi(t) \longmapsto g \cdot ( \chi(t)) ) 
   = \frac{(\varphi_g(t))^{w-2}}{k_1^{\frac{w}{2}-1}}\chi(t^g)
    \in \C[t]_{{\rm deg}=w-2}. 
\end{align}
The assignment $P_f$ for $f$ in (\ref{eq:assign-PP-holCase}) is 
regarded as a 1-cocycle 
\begin{align}
  P_f \in Z^1(G, \C[t]_{{\rm deg}=w-2})
\end{align}
in the sense of group cohomology. The integration constant terms 
$Q(t)$ to be added to the Eichler integral 
(cf (\ref{eq:add-intg-cnst-2EichlerI})) gives rise to 
the ambiguity to be added to $P_g(t, f)$ as 
in (\ref{eq:PP+cobndry=anal.contn.violatn}). This additional 
ambiguity is interpreted as the coboundary from 0-cochains 
in group cohomology, so a cusp form $f$ of weight $w$ determines 
an element in the group cohomology 
\begin{align}
  [P_f] \in H^1(G, \C[t]_{{\rm deg}=w-2}). 
\end{align}

In this article, we had to deal with modular forms $f_*$ that may 
have a pole in the interior of the complex upper half plane. 
Now, the group $G=[\Gamma_S]$ such as ${\rm PSL}_2\Z$ 
(for $\Lambda_S = \vev{+2}$) and a finite index subgroup of $\Gamma_0(k)_+$ 
(for $\vev{+2k}$) is replaced by 
\begin{align}
 \pi_1 \left( (D(\widetilde{\Lambda}_S) \backslash X_{\rm singl})/[\Gamma_S], 
     \; [t_0] \right),
   \label{eq:pi1-generic-part-modSp}
\end{align}
and the period polynomials allow us to think of an assignment 
\begin{align}
  P_{f_*} : \pi_1 \ni \gamma_{\tilde{g}} \longmapsto
     P_{\tilde{g}}(t, f_*) \in \C[t]_{{\rm deg}=w-2},  
\end{align}
where we use an abbreviated notation $\pi_1$ for the group 
in (\ref{eq:pi1-generic-part-modSp}). 
The condition (\ref{eq:monodrm-product-compatible-PP}) implies 
that this assignment is a 1-cocycle of the group $\pi_1$ 
(be aware that we use the standard loop composition law in the 
group $\pi_1$ here, whereas the composition of the duality transformation 
$\tilde{g}_a\cdot \tilde{g}_b$ corresponds 
to (\ref{eq:path-composition-law-2})). 

It has been explained in this article that the modular form $f_*$ 
is determined uniquely by the indices (elliptic genus) $\{ \Phi_\gamma \}$. 
This fact means that the classification of branches of moduli space by 
(implicit $\Lambda_S$ and) $\{\Phi_\gamma \}$---referred to as the coarse 
classification in \cite{EW19}---is equivalent to classification 
by the Eichler cohomology 
\begin{align}
  P_{f_*} + B^1(\pi_1, \; \C[t]_{{\rm deg}=4}) \in H^1(\pi_1, \C[t]_{{\rm deg}=4}). 
\end{align}

On the other hand, we also studied the monodromy representations 
that are formulated by using the BPS dyon spectrum of branches 
of moduli space. The diagonal part of the monodromy matrices 
is in $[\Gamma_S]$, and is determined by the lattice $\Lambda_S$;  
the off-diagonal part (the matrices $\Lambda_{\tilde{g}}$)  
carry more detailed information of the branches of the moduli space. 
We have seen in this article that one can think of the assignment 
\begin{align}
 \Lambda:
    \pi_1 \ni \gamma_{\tilde{g}} \longmapsto \Lambda_{\tilde{g}} + \Z \widetilde{C}
   \in {\rm Sym}^2(\Z^{\oplus (2+\rho)})/\Z\widetilde{C}. 
\end{align}
The group $[\Gamma_S]$ acts on the abelian group 
${\rm Sym}^2(\Z^{\oplus(2+\rho)})/\Z\widetilde{C}$ as 
\begin{align}
 (\Lambda + \Z\widetilde{C}) \longmapsto 
     g^T \cdot (\Lambda + \Z \widetilde{C}) \cdot g. 
\end{align}
The condition (\ref{eq:monodrm-product-compatible-matrix})
implies that the assignment $\Lambda$ above is a 1-cocycle 
of the group $\pi_1$ (be aware of the composition law, once again). 

There is no unique choice of a symplectic frame at the base point $t_0$. 
The ambiguity in the choice of a frame discussed in 
footnote \ref{fn:symp-frame-change} is parametrized by just one matrix 
\begin{align}
  \Lambda_0 \in {\rm Sym}^2(\Z^{\oplus (2+\rho)}), \qquad 
   (\Lambda_0)_{00} = (\Lambda_0)_{0\sharp} = (\Lambda_0)_{\sharp \sharp} = 0. 
   \label{eq:restrict-symp-frame-choice}
\end{align}
This can be seen as a 0-cochain, and the change in the matrix 
$\Lambda_{\tilde{g}}$ under the change in the choice of a symplectic frame 
at the base point is regarded as the coboundary of a 0-cochain 
of the form above. So, one may say that the main text of this article 
classified branches of the Heterotic--IIA moduli space by 
using 
\begin{align}
 \Lambda + B^1(\pi_1, \; {\rm Sym}^2(\Z^{\oplus (2+\rho)}
    )_{{\rm condt'n}(\ref{eq:restrict-symp-frame-choice})}/\Z\widetilde{C}). 
  \label{eq:dyon-spectrum-monodromy-inv}
\end{align}

Now, we are ready to describe a role played by the invariants 
$\{ \Psi_\gamma \}$ for finer classification in \cite{EW19}, using 
the language of group cohomology. For branches of the moduli space 
that share the same $\{ \Phi_\gamma \}$, and hence the same $f_*$, 
there are still distinct dyon spectra and their monodromy 
characterized by (\ref{eq:dyon-spectrum-monodromy-inv}). 
Such branches distinguished by (\ref{eq:dyon-spectrum-monodromy-inv})
are one-to-one with the set 
\begin{align}
  \left[ 
 \left( P_{f_*} + B^1(\pi_1, \; \C[t]_{{\rm deg}=4}) \right) \cap 
    Z^1(\pi_1, \; \Z[t]_{{\rm deg}=4}) \right] / 
  B^1(\pi_1, \; \Z[t]_{{\rm deg}=4, \; {\rm condt'n}(\ref{eq:restrict-symp-frame-choice})});
  \label{eq:classificatn-set-by-dyon}
\end{align}
symmetric matrices $\Lambda$ (mod $+\Z\widetilde{C}$) have been converted 
to polynomials through $\mho^T(t) \cdot \Lambda \cdot \mho(t)$ 
as we have done in the main text. The set (\ref{eq:classificatn-set-by-dyon})
is not necessarily of single element (e.g.\cite{MV-2, BW-16}); the 
classification invariant $\{ \Psi_\gamma \}$ was introduced 
in \cite{EW19} so that elements in this set are distinguished. 
Presumably the authors of \cite{AFGNT, AP} anticipated classification 
by something like (\ref{eq:classificatn-set-by-dyon}); to get it done, 
we had to deal with integers (rather than $\Q$, $\R$ or $\C$) 
by paying enough attention to normalizations in this article; 
direct computation of period polynomials made a systematic study of 
the set (\ref{eq:classificatn-set-by-dyon}) possible.

\end{document}